\documentclass[aps,prl,amsmath,amssymb,superscriptaddress]{revtex4-1}
\usepackage{dcolumn}
\usepackage{amsmath}
\usepackage{amssymb}
\usepackage{graphicx}
\usepackage{bm}
\usepackage{epstopdf}

\begin{document}
\title{Transitions from a Kondo-like diamagnetic insulator into a modulated ferromagnetic metal in  $\bm{\mathrm{FeGa}_{3-y}\mathrm{Ge}_y}$}

\affiliation{Department of Chemistry, Graduate School of Science, Kyoto University, Kyoto 606-8502, Japan}
\affiliation{School of Chemistry and Chemical Engineering, Shanghai Jiao Tong University, Shanghai, 20040, China}
\affiliation{Graduate School of Advanced Sciences of Matter, Hiroshima University, Higashi-Hiroshima, 739-8530, Japan}
\affiliation{Centro de Ci\^{e}ncias Naturais e Humanas, Universidade Federal do ABC, Santo Andr\^{e}-SP, 09210-580, Brazil}
\affiliation{Research Center for Low Temperature and Materials Sciences, Kyoto University, Kyoto 606-8502, Japan}

\author{Yao Zhang}
\email[]{yao.ns@sjtu.edu.cn}
\affiliation{Department of Chemistry, Graduate School of Science, Kyoto University, Kyoto 606-8502, Japan}
\affiliation{School of Chemistry and Chemical Engineering, Shanghai Jiao Tong University, Shanghai, 20040, China}

\author{Masaki Imai}
\affiliation{Department of Chemistry, Graduate School of Science, Kyoto University, Kyoto 606-8502, Japan}

\author{Chishiro Michioka}
\affiliation{Department of Chemistry, Graduate School of Science, Kyoto University, Kyoto 606-8502, Japan}

\author{Yuta Hadano}
\affiliation{Graduate School of Advanced Sciences of Matter, Hiroshima University, Higashi-Hiroshima, 739-8530, Japan}

\author{Marcos A. Avila}
\affiliation{Centro de Ci\^{e}ncias Naturais e Humanas, Universidade Federal do ABC, Santo Andr\^{e}-SP, 09210-580, Brazil}

\author{Toshiro Takabatake}
\affiliation{Graduate School of Advanced Sciences of Matter, Hiroshima University, Higashi-Hiroshima, 739-8530, Japan}

\author{Kazuyoshi Yoshimura}
\email[]{kyhv@kuchem.kyoto-u.ac.jp}
\affiliation{Department of Chemistry, Graduate School of Science, Kyoto University, Kyoto 606-8502, Japan}
\affiliation{Research Center for Low Temperature and Materials Sciences, Kyoto University, Kyoto 606-8502, Japan}

\date{June 5, 2017}

\begin{abstract}
One initial and essential question of magnetism is whether the magnetic properties of a material are governed by localized moments or itinerant electrons. Here we expose the case for the weakly ferromagnetic system FeGa$_{3-y}$Ge$_y$ wherein these two opposite models are reconciled, such that the magnetic susceptibility is quantitatively explained by taking into account the effects of spin-spin correlation. With the electron doping introduced by Ge substitution, the diamagnetic insulating parent compound FeGa$_3$ becomes a paramagnetic metal as early as at $ y=0.01 $, and turns into a weakly ferromagnetic metal around the quantum critical point $ y=0.15 $. Within the ferromagnetic regime of FeGa$_{3-y}$Ge$_y$, the magnetic properties are of a weakly itinerant ferromagnetic nature, located in the intermediate regime between the localized and the itinerant dominance. Our analysis implies a potential universality for all itinerant-electron ferromagnets.
\end{abstract}

\maketitle
 
The studies on spin-orbit couplings, the Bir-Aronov-Pikus mechanism and related spin relaxation mechanisms, Dzyaloshinsky-Moriya (D-M) interactions etc., have recently revealed intriguing phenomena in many fascinating research areas, such as spintronics \cite{Zutic2004,Ohno1999,Wolf2001,Sun2003}, skyrmions \cite{Sondhi93,Senthil2004,Rossler06},  spin caloritronics \cite{Bauer10,Bauer12}. Understanding the correlations of microscopic alignment of spin moments, i.e. the ingredients of magnetic mechanisms, is crucial to academic studies as well as technical applications. Spin fluctuations in many-body systems are of such importance, as it leads to the formation of \textquotedblleft strange metals\textquotedblright, the non-Fermi liquid, extended to high temperature \cite{Pfleiderer2008,Smith2008}. In the cuprate and iron-based superconductors, the essential pairing interaction is proved to be mediated by the spin fluctuations as a common thread in the unconventional superconductors \cite{Scalapino2012,Wu2017}. 

In the field of magnetism, one open and important issue is to establish a unified model for itinerant ferromagnets.  Well-established theories are restricted to two narrow extremes,  i.e., the localized and itinerant-electron regimes. Though extensive effort has been made to elucidate the magnetic properties in the intermediate range of these two opposite extremes \cite{Falicov1969,Izuyama1963,Lederer1966}, a successful theory has proved to be elusive. A recent picture of the hybrid nature of localized moments and itinerant electrons was explored in several systems, on which the hybrid model was proposed in a two-band approximation to illustrate the magnetism and in some cases the origin of superconductivity \cite{Vojta2009,Xu2009,Paglione2010,Dai2012,Kou2009,you2011l,You2011b,Hu2012,Dai2009}. The self-consistent renormalization (SCR) theory of spin fluctuations and related theories successfully approach the localized regime based on the itiernant picture as an intermediate mechanism in the one-band model by mediating the magnetic momentum of itinerant electrons in terms of wave-number-dependent spin fluctuation and generalized dynamical fluctuations \cite{Moriya1985, Moriya1978}. Despite this, however, a unified dynamical theory is still being debated paticularly due to the limitated diversity for further study, as well as the difficulty in reconciling these two polar extremes \cite{Svanidze2015}.

The heavy-fermion Kondo insulators provide a good platform to explore the physical properties including magnetic ordering due to the coupling of the charge dynamics to the component ordering associated with its related fluctuations during a metal-insulator transition. The Kondo-insulator-like semiconductor, FeGa$_3$, which has a larger pseudogap compared to the typical Kondo insulators, is such an ideal system owing to its expected valence admixture \cite{Varma1994,Tomczak2012}. The energy gap of FeGa$_3$ is about 0.4 eV, and its pseudogap is formed by the strong hybridization between the $3d$ band of Fe and $4p$ band of Ga \cite{Arita2011,Tsujii2008,Haussermann2002}. No magnetic ordering is detected in FeGa$_3$ by $^{57}$Fe M\"{o}ssbauer experiments \cite{Tsujii2008}.
The nonmagnetic FeGa$_3$ is reminiscent of another Fe-based Kondo insulator: FeSi, which has drawn attention for decades.
FeSi undergoes a first order transition and turns into a ferromagnetic metal at the critical composition of FeSi$ _{0.75} $Ge$ _{0.25}$ with modest electron doping by Ge  \cite{Yeo2003}. The Co substitution introduces strong magnetic resistivity, anomalous Hall effect and a chiral magnetic nature into Fe$ _{1-x} $Co$ _{x} $Si system. The interesting magnetic properties induced by the electron doping such as spiral magnetic structure and reentrant spin glass behaviors are represented by the D-M interaction, conventional isotropic exchange (J), anisotropic exchanges, the Zeeman interaction under applied field and cubic anisotropy effects \cite{Chattopadhyay2002,Grigoriev2007a}. 
In this report, we follow the phase transitions by increasing the Ge doping of FeGa$_{3-y}$Ge$_y$ from a nonmagnetic Kondo-like insulator to a paramagnetic metal and finally to a weakly ferromagnetic metal, through all of which the magnetic properties are significantly affected. In the ferromagnetic region, the magnetic ordering of FeGa$_{3-y}$Ge$_y$ is shown to be located in a crossover between the localized model and itinerant regimes, and the localized/itinerant character in electrons can be enhanced/diminished by the filling control. A quantitative explanation of the magnetic properties is achieved by reconciling the localized model with the itinerant model by mainly taking into account the effects of temperature dependent spin fluctuations in general. Spin fluctuations may play a key role in approaching a unified theory for localized and itinerant ferromagnets, and FeGa$_{3-y}$Ge$_y$ seems to be one of the best candidates for probing the unified theory for itinerant magnetism.
\\

\section{Results}

\begin{figure}[tbp] 
	\centering
	\includegraphics[width=0.7\linewidth,keepaspectratio]{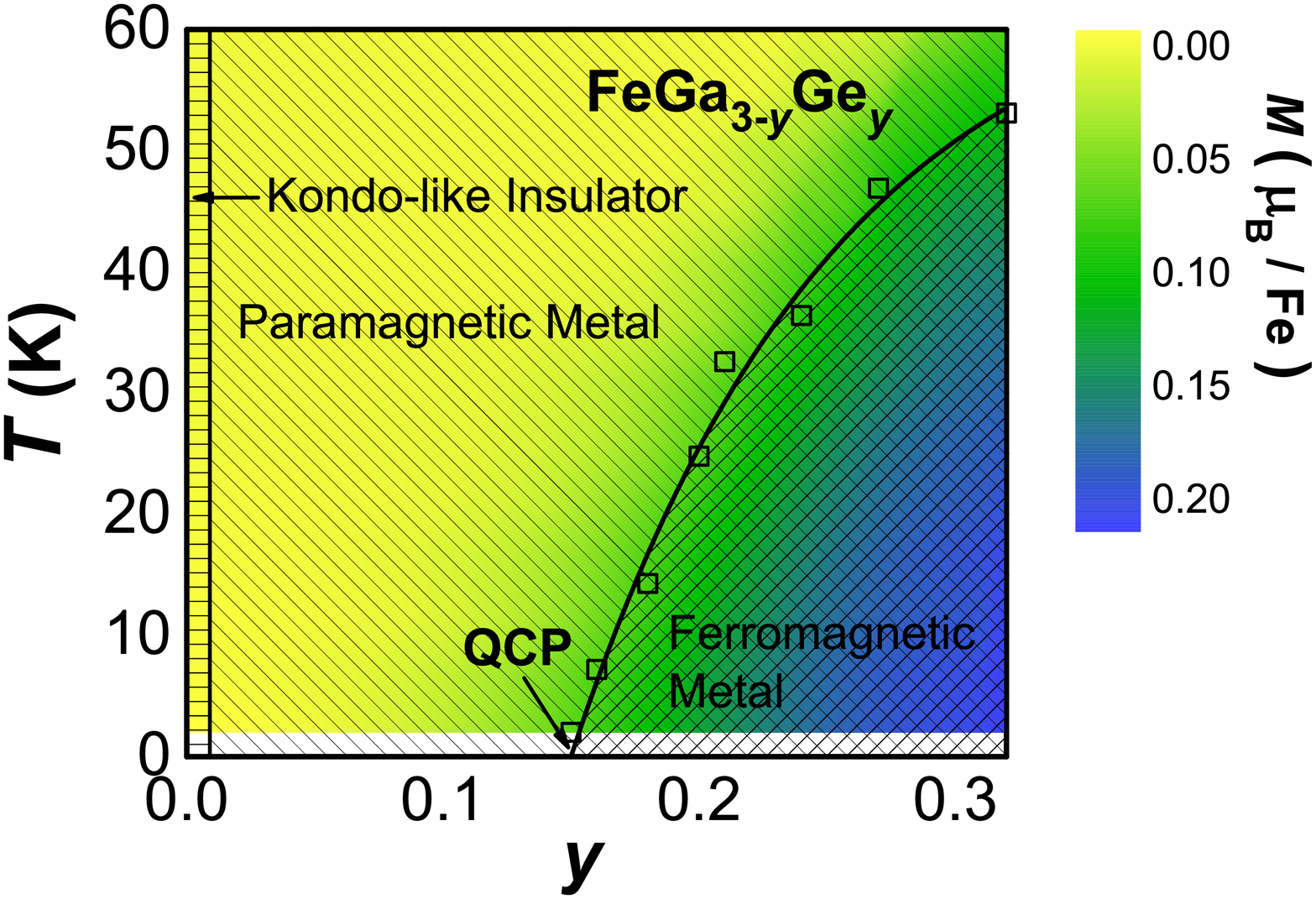}
	\caption{\textbf{Phase diagram of FeGa$_{3-y}$Ge$_y$.}  Open squares represent the ferromagnetic transition temperature $ T_{\mathrm{C}} $ and the bold arrow shows the quantum critical point $ (\textrm{at } y=0.15) $. Solid line at $ y=0.01 $ corresponds to critical transition edge between the Kondo-like insulator and the paramagnetic metal. Color scale represents magnetization of FeGa$_{3-y}$Ge$_y$ measured at $ H = 1 \mathrm{T}$  at various temperatures.
		\label{fig:P-D}}  
\end{figure}

\textbf{Phase transitions.} Figure \ref{fig:P-D} displays the phase diagram of FeGa$_{3-y}$Ge$_y$ obtained on the basis of the magnetic and transport measurements (see supplement Figs. S1 and S2). The open squares indicate the ferromagnetic transition temperatures $T_{\mathrm{C}}$ and the bold arrow  indicates the quantum critical point (QCP). Increasing Ge substitution for Ga results in FeGa$_{3-y}$Ge$_y$ turning from a diamagnetic insulator into a paramagnetic metal, and eventually into a weakly ferromagnetic metal. The magnetization under $H$ = 1 T is shown by the color scale. The spontaneous magnetic moment $P_{\mathrm{S}}$ of FeGa$_{3-y}$Ge$_y$ is estimated in Arrott plots in Fig. \ref{fig:M2-M4} (and Fig. S4) by extrapolating the linear relation, and the values are summarized in Table 1. The relatively small magnitude indicates a weakly ferromagnetic nature. The critical point of electric and magnetic transitions for FeSi$_{1-x}$Ge$_{x}$ occurred at  $x = 0.25$, and the energy gap of FeSi is about one tenth of that of FeGa$_3$, the Ge substitution in FeGa$_3$ considerably affects the electronic state in FeGa$_{3-y}$Ge$_y$.

\begin{figure}[tbp] 
	\centering
	\includegraphics[width=0.6\linewidth,keepaspectratio]{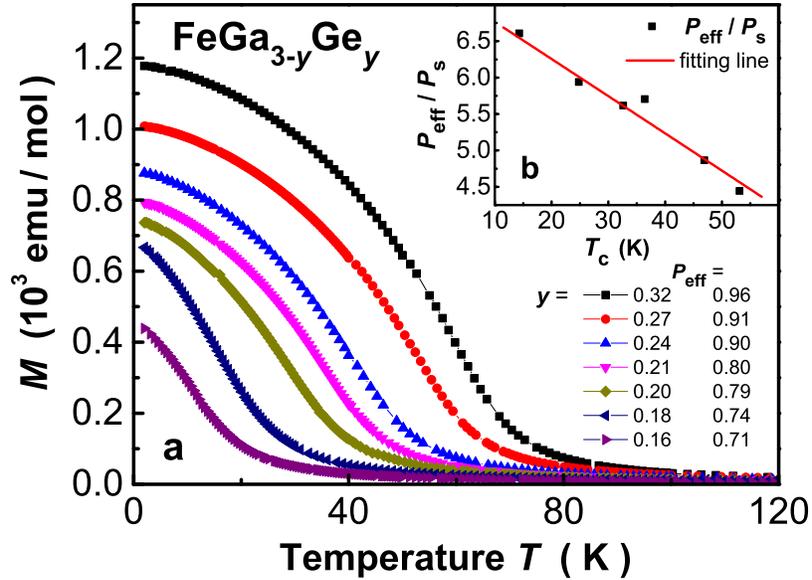}
	\caption{\textbf{Temperature dependent magnetization, and linear relation bewteen \boldmath{$ P_{\mathrm{ eff}}/P_{\mathrm{S}} $}  and \boldmath{$ T_{\mathrm{C}} $.}} \textbf{(a)} $ T $ dependence of $ M $ of ferromagnetic FeGa$_{3-y}$Ge$_y$ at $ H $ = 1 T. $ P_{\mathrm{ eff}} $ are obtained by fitting the Curie-Weiss law using data at high temperature region. \textbf{(b)} $ P_{\mathrm{eff}}/P_{\mathrm{S}} $  versus $ T_{\mathrm{C}} $ plots. Solid line is the fitting line: $ P_{\mathrm{eff}}/P_{\mathrm{S}}=-1/20~T_{\mathrm{C}}^{1.005} +c$.
		\label{fig:MT-i}}  
\end{figure}

\textbf{Magnetic properties and magnetic orderings.} The magnetization $ M $ versus temperature $ T $, and  $ P_{\mathrm{eff}}/P_{\mathrm{S}} $ versus $ T_{\mathrm{C}} $  are shown for $ y\geq0.16 $ in Fig. \ref{fig:MT-i}a and Fig. \ref{fig:MT-i}b, respectively. The magnetization increases sharply with decreasing temperature in the low-temperature region, following the typical ferromagnetic behavior. Plots follow the Curie-Weiss (CW) law in the high-temperature region, (see Fig. \ref{fig:CE} and Supplementary Fig. 6). The effective magnetic moment $ P_{\mathrm{eff}} $ for various $ y $  are estimated from $ M $ versus $ T $ plots above 150 K. $ P_{\mathrm{eff}} $ displays a relatively weak $ y $ dependence as an itinerant ferromagnet. In Fig. \ref{fig:MT-i}b, the relation of $ P_{\mathrm{eff}} $/$ P_{\mathrm{S}} $ and $ T_{\mathrm{C}} $ can be fitted by a nearly linear function $ P_{\mathrm{eff}}/P_{\mathrm{S}}  = -a \hspace{0.2mm} T_{\mathrm{C}}^{(1+\beta)}+c $, with $ a=-1/20$, $ \beta \sim 0.005 $, and $ c=7.27 $.

\begin{figure*}[t] 
	\centering
	\begin{tabular}{c@{\hskip 0.7in}c}
		\includegraphics[width=0.35\textwidth,keepaspectratio]{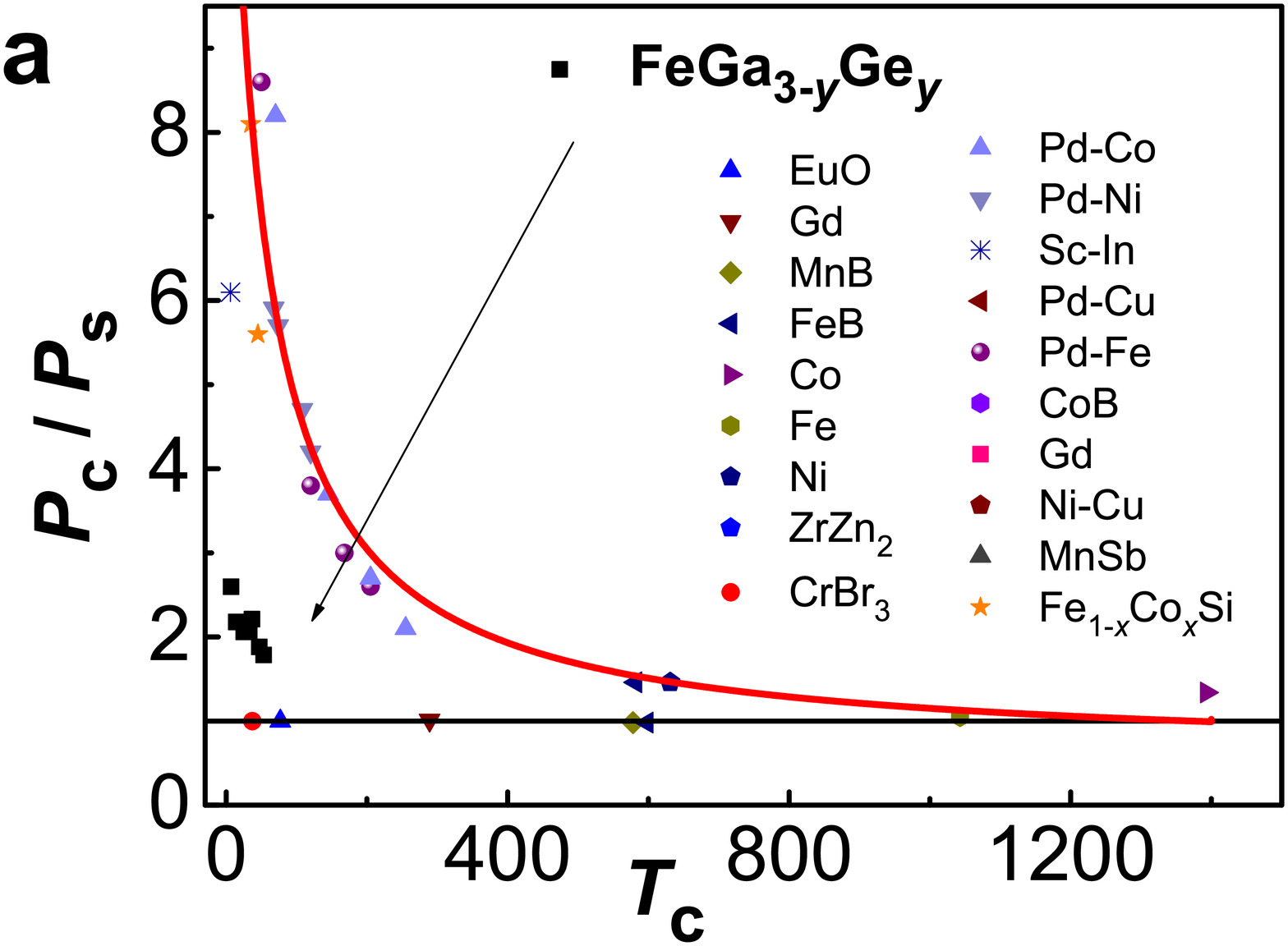}& \includegraphics[width=0.35\textwidth,keepaspectratio]{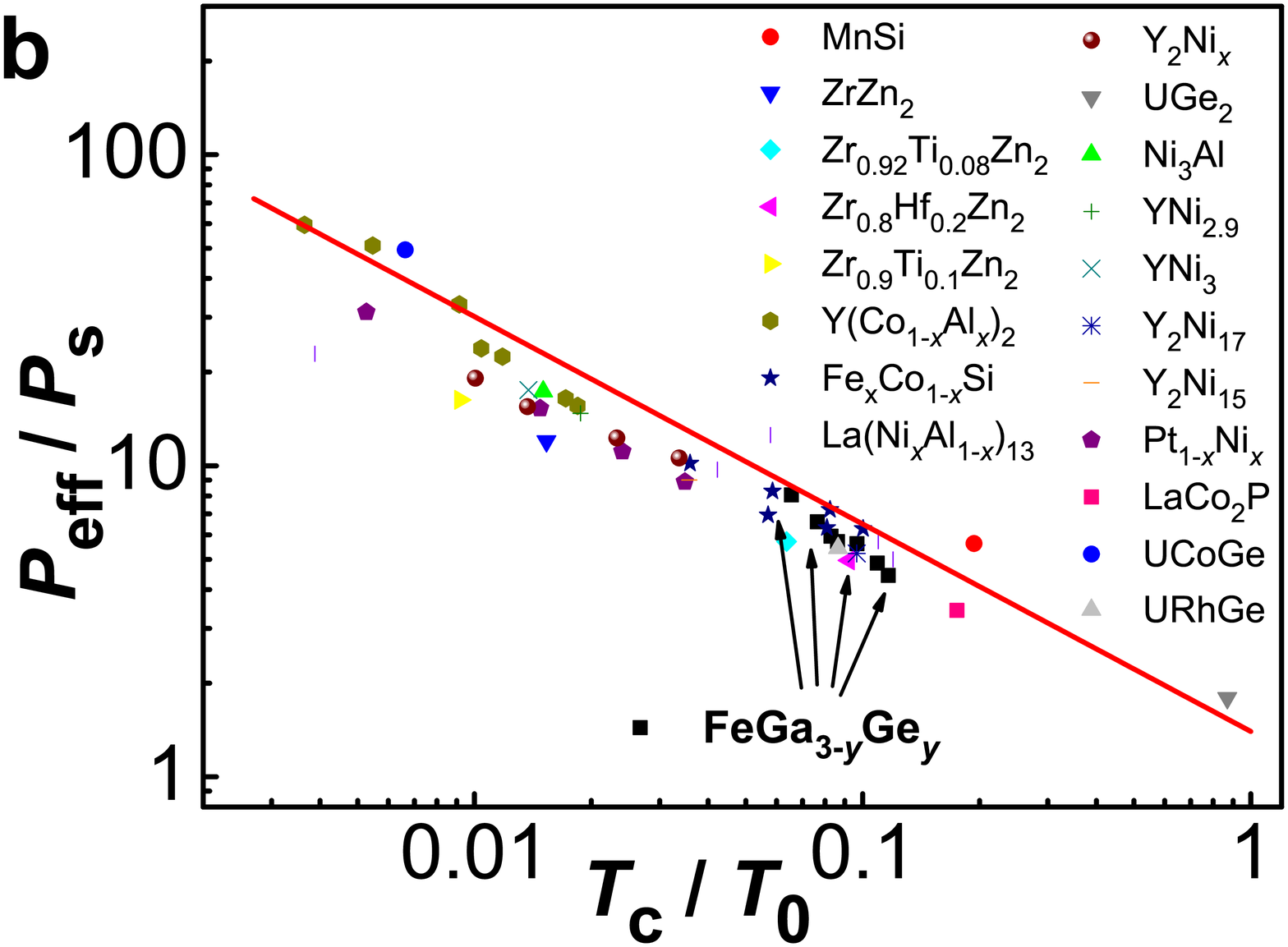}\\
	\end{tabular}
	\caption {
		\textbf{Rhodes-Wohlfarth plot and Deguchi-Takahashi plot.} 
		{\fontfamily{phvr}\selectfont
			$ P_{\mathrm{C}}/P_{\mathrm{S}} $ versus $ T_{\mathrm{C}} $ plot and $ P_{\mathrm{eff}}/P_{\mathrm{S}}$ versus $ T_\mathrm{C}/T_0 $ plot for FeGa$_{3-y}$Ge$_y$ and various ferromagnets, as ({\fontfamily{phv}\selectfont a}) and ({\fontfamily{phv}\selectfont b}) respectively. Data are reproduced from Refs. \cite{Rhodes1963,Wohlfarth1978,Bloch1975,Ogawa1968,Beille1974,Yoshimura1987,Deboer1969,Ogawa1976,Shimizu1990,Nakabayashi1992,Fujita1995,Reehuis1994}.
			({\fontfamily{phv}\selectfont a}) Parameters of FeGa$_{3-y}$Ge$_y$ do not follow the universal line, and $ P_{\mathrm{C}}/P_{\mathrm{S}} $ of  FeGa$_{3-y}$Ge$_y$ is relatively small compared with other ferromangets with same magnitude of $ T_{\mathrm{C}} $. ({\fontfamily{phv}\selectfont b}) Red straight line represents Takahashi's theoretical line, $ P_{\mathrm{eff}}/P_{\mathrm{S}} =1.4\times(T_\mathrm{C}/T_0 )^{-2/3}  $, which roughly describes FeGa$_{3-y}$Ge$_y$.   }
		\label{fig:RW-DT}}  
\end{figure*}

\begin{figure*}[bp]
	\centering
	\begin{tabular}{cc@{\hskip 0.6in}cc}
		\includegraphics[width=0.22\linewidth,keepaspectratio]{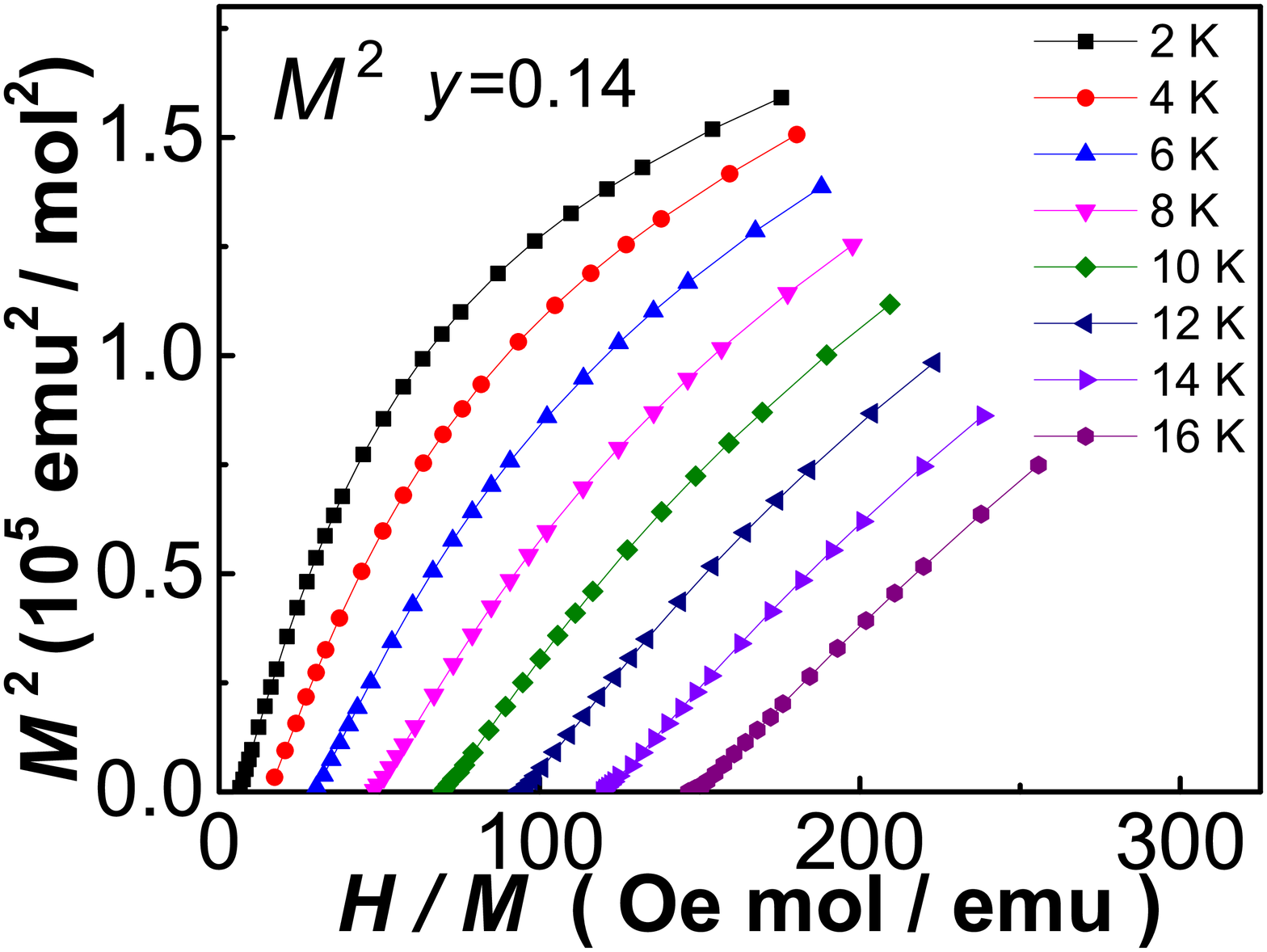}& \includegraphics[width=0.22\linewidth,keepaspectratio]{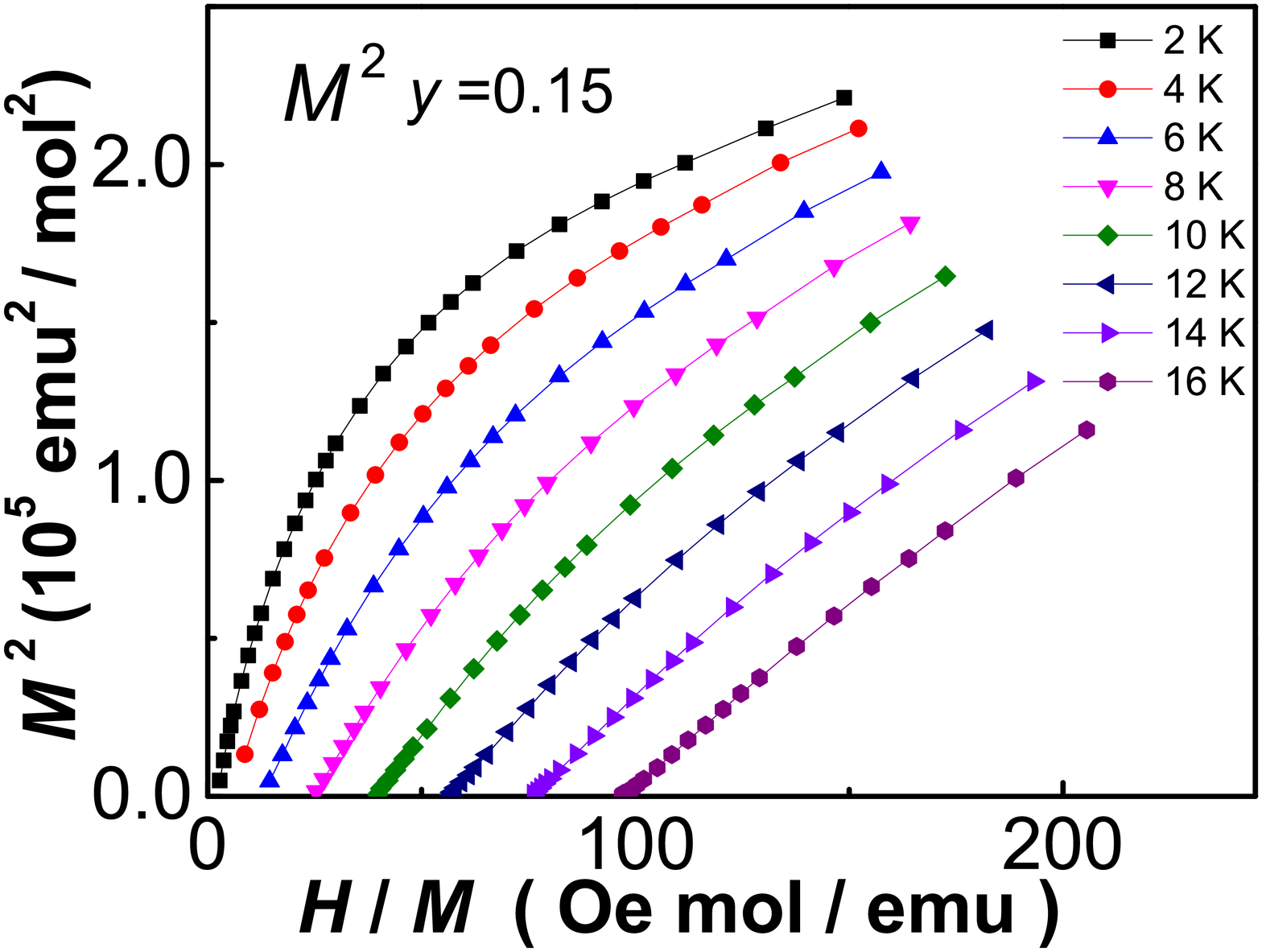}&\includegraphics[width=0.22\linewidth,keepaspectratio]{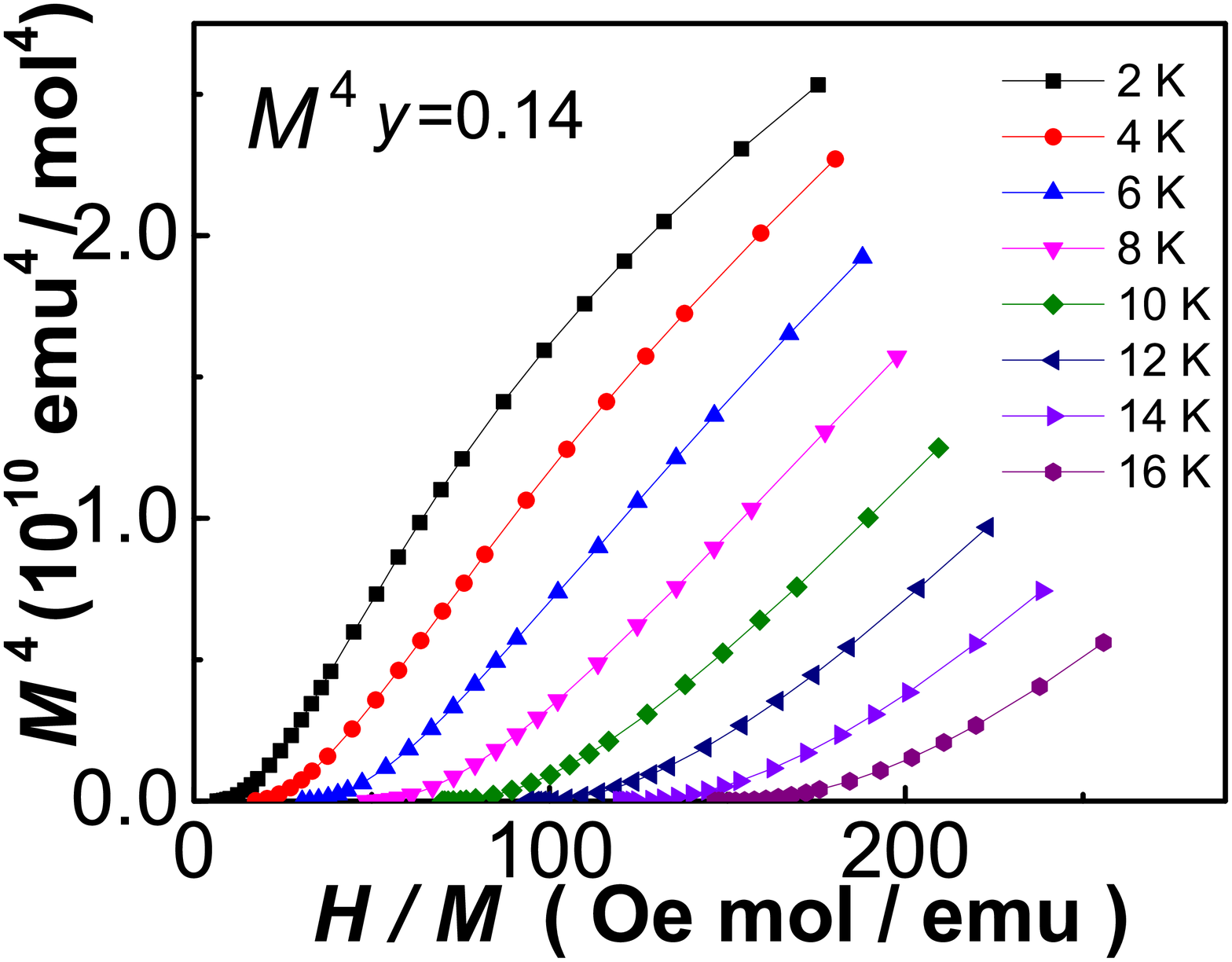}& \includegraphics[width=0.22\linewidth,keepaspectratio]{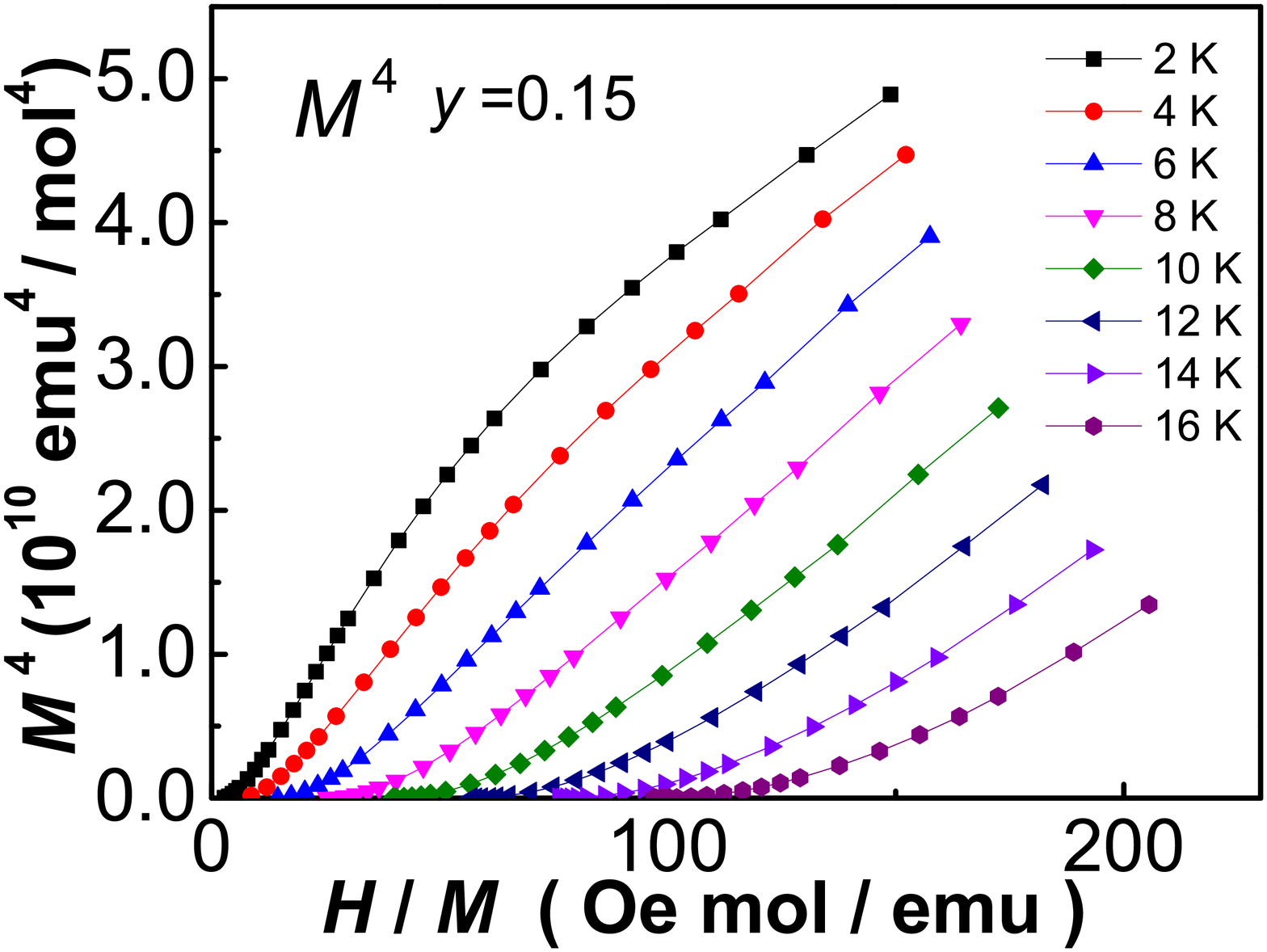}\\
		\includegraphics[width=0.22\linewidth,keepaspectratio]{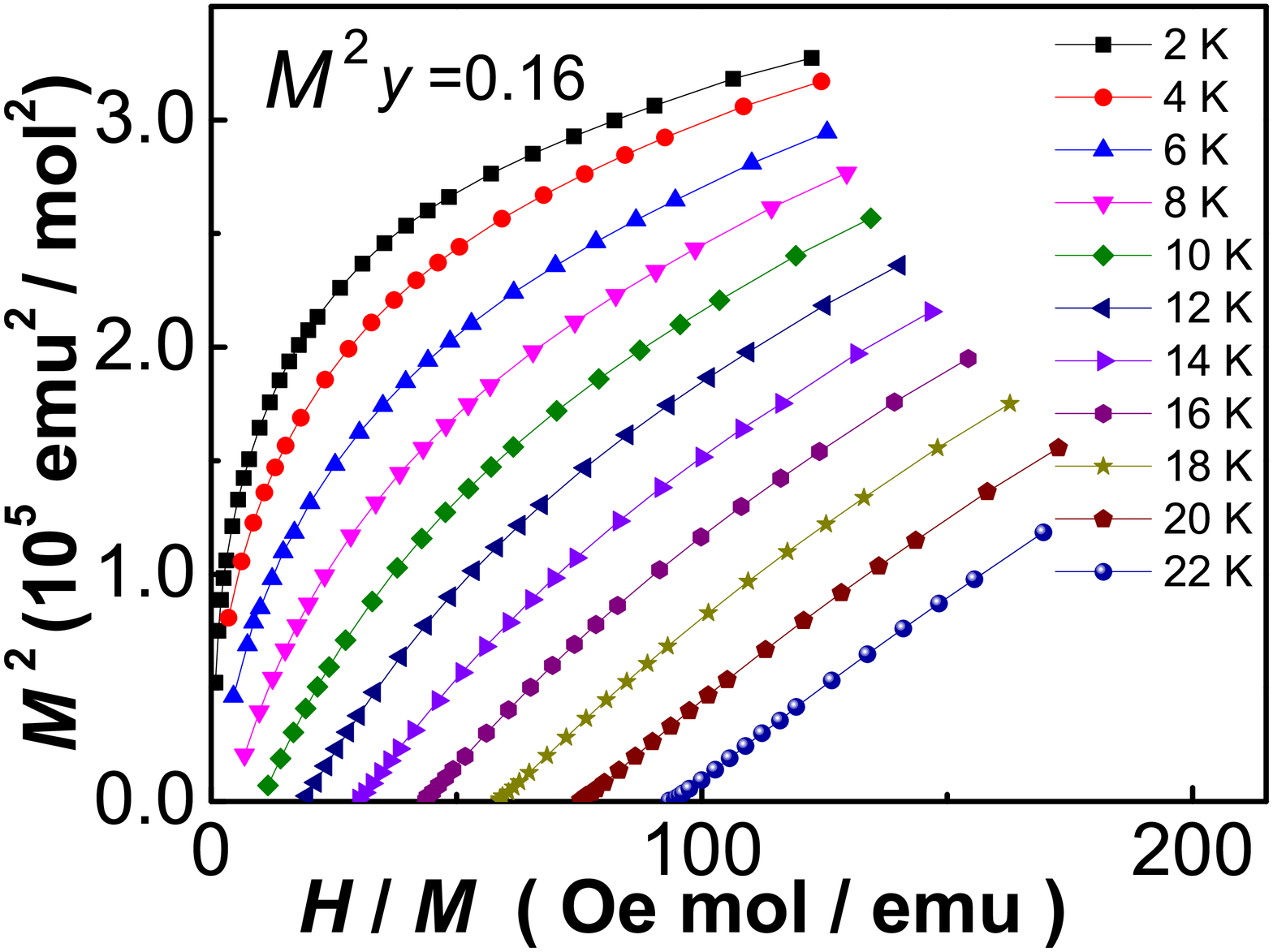}& \includegraphics[width=0.22\linewidth,keepaspectratio]{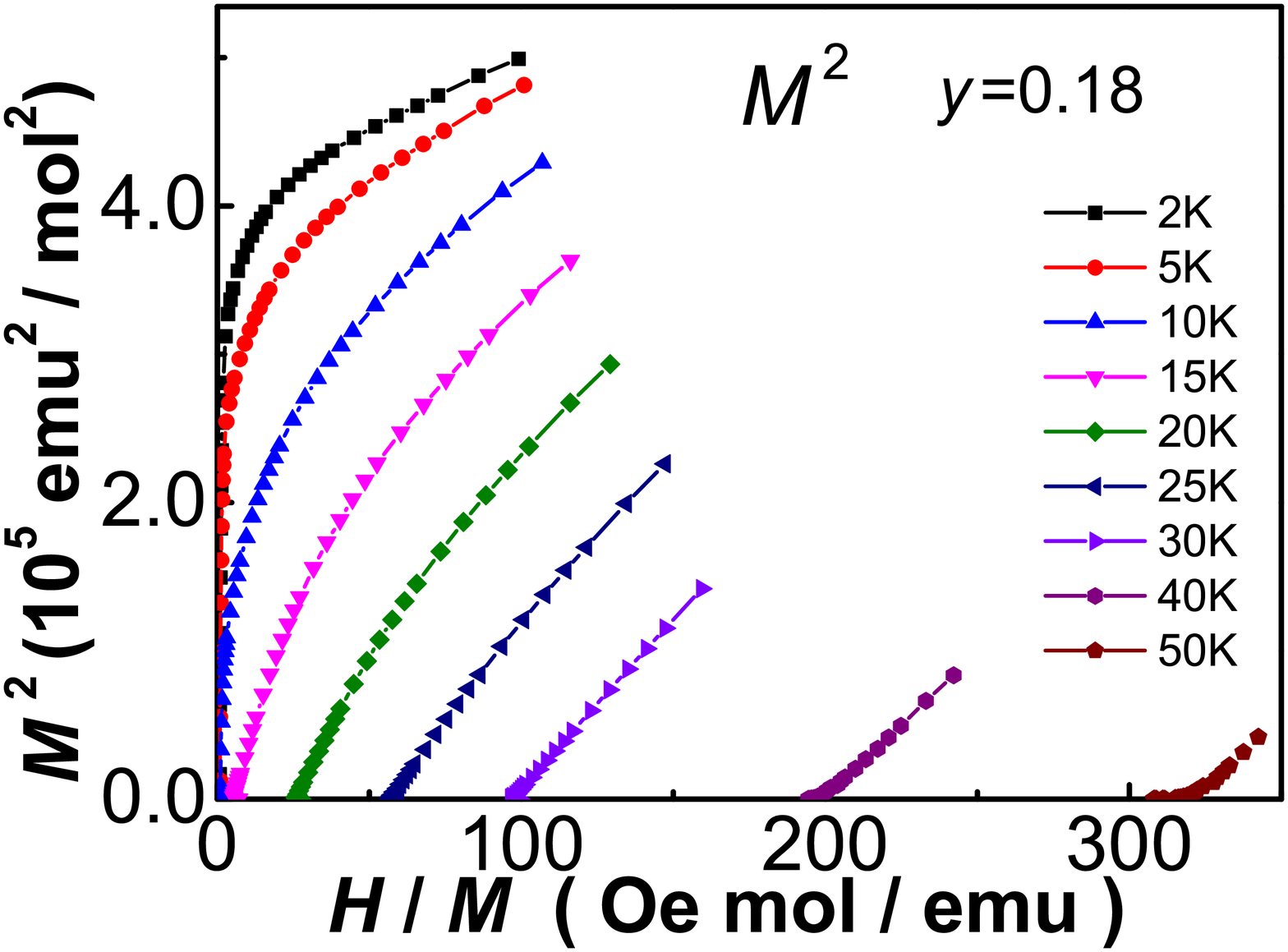}&\includegraphics[width=0.22\linewidth,keepaspectratio]{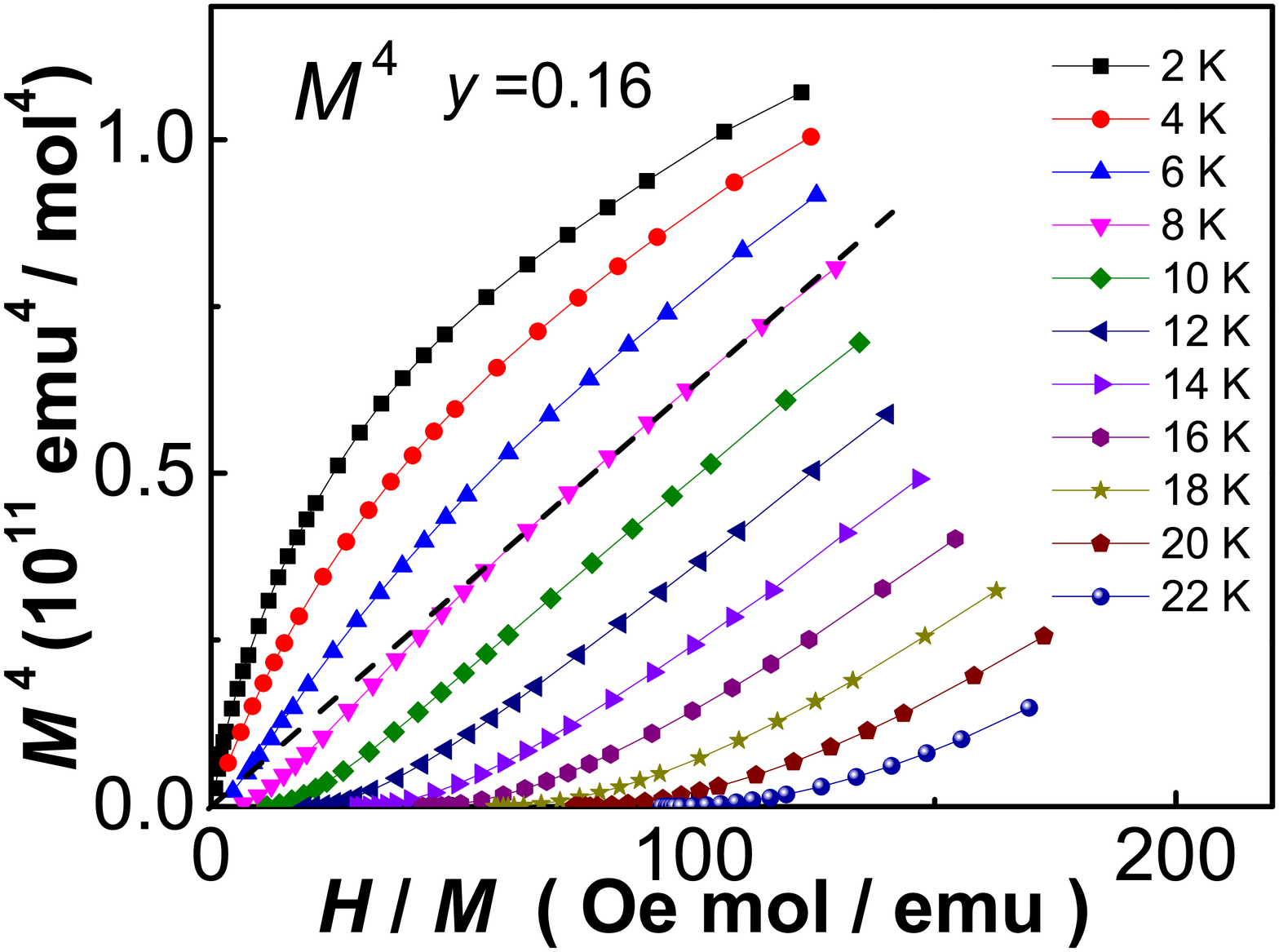}& \includegraphics[width=0.22\linewidth,keepaspectratio]{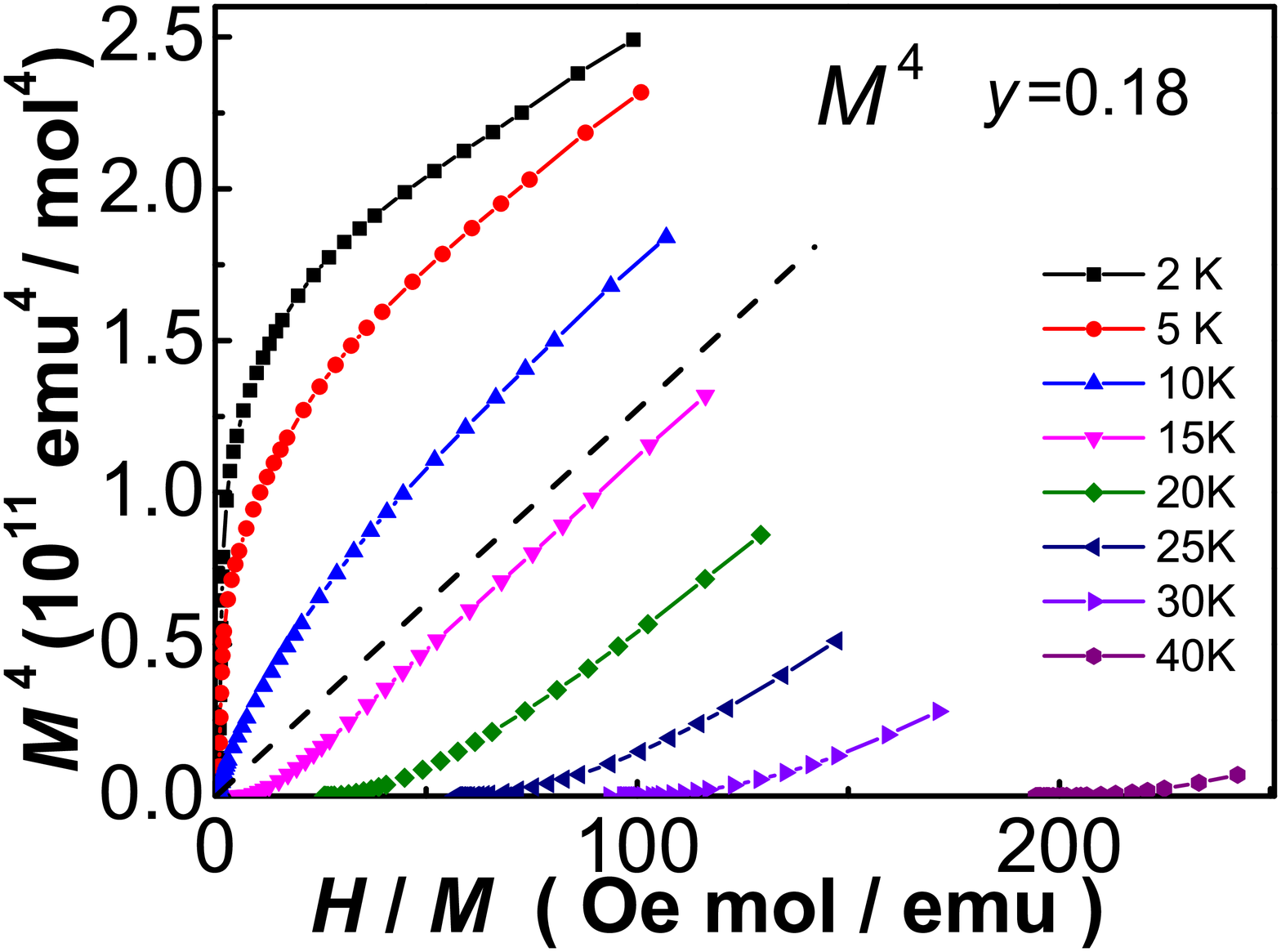}\\
		\includegraphics[width=0.22\linewidth,keepaspectratio]{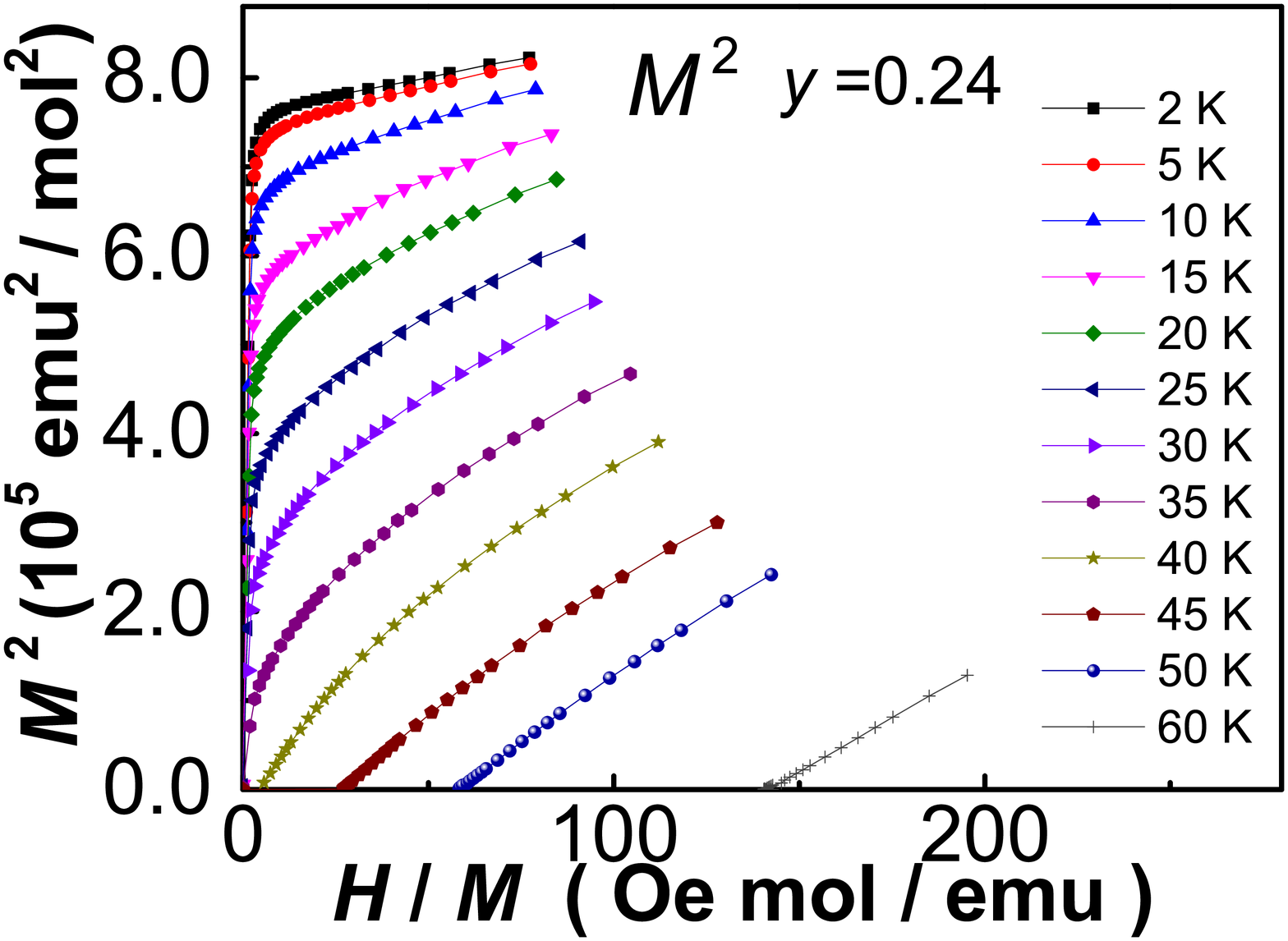}&  
		\includegraphics[width=0.22\linewidth,keepaspectratio]{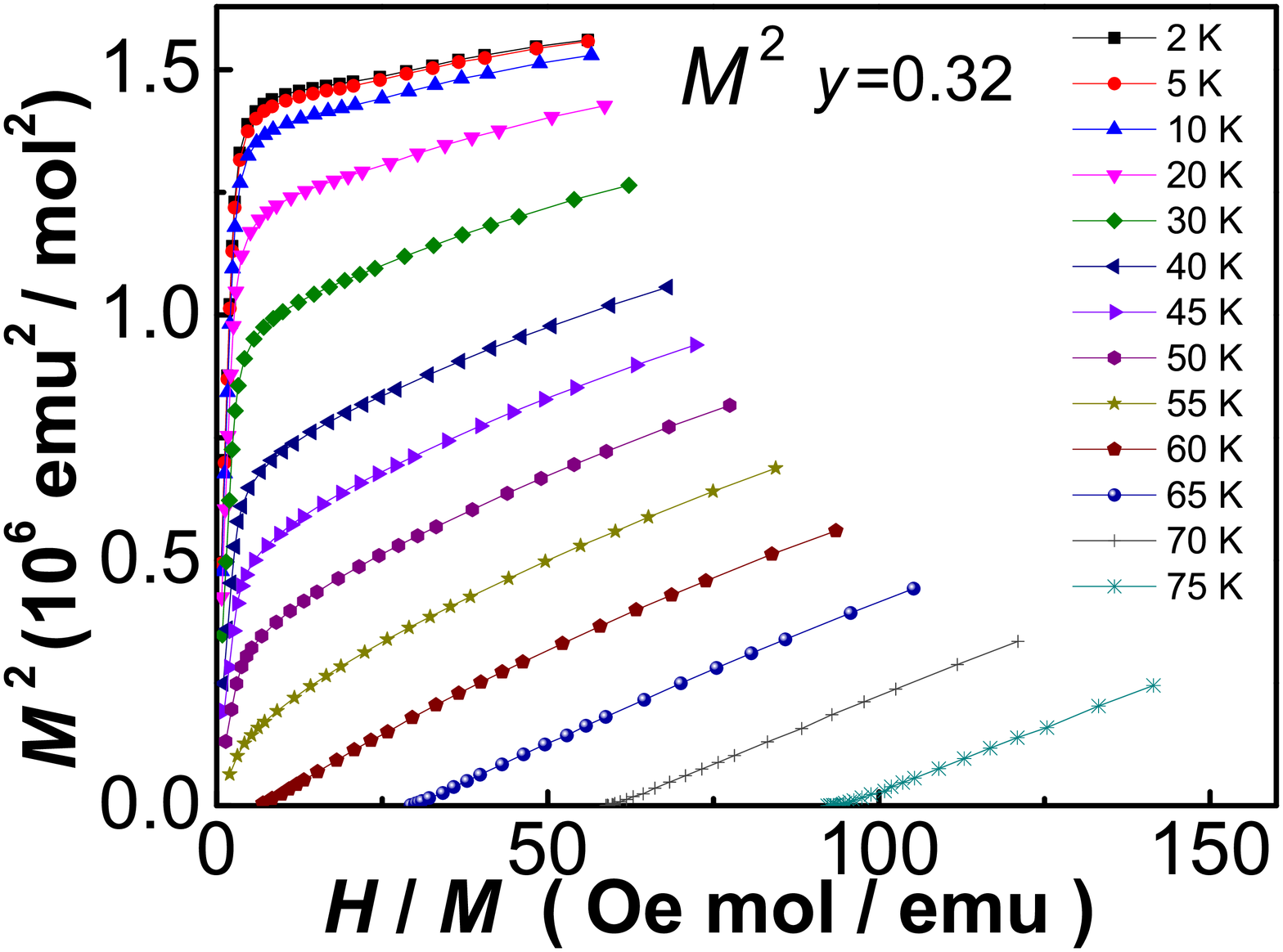}&\includegraphics[width=0.22\linewidth,keepaspectratio]{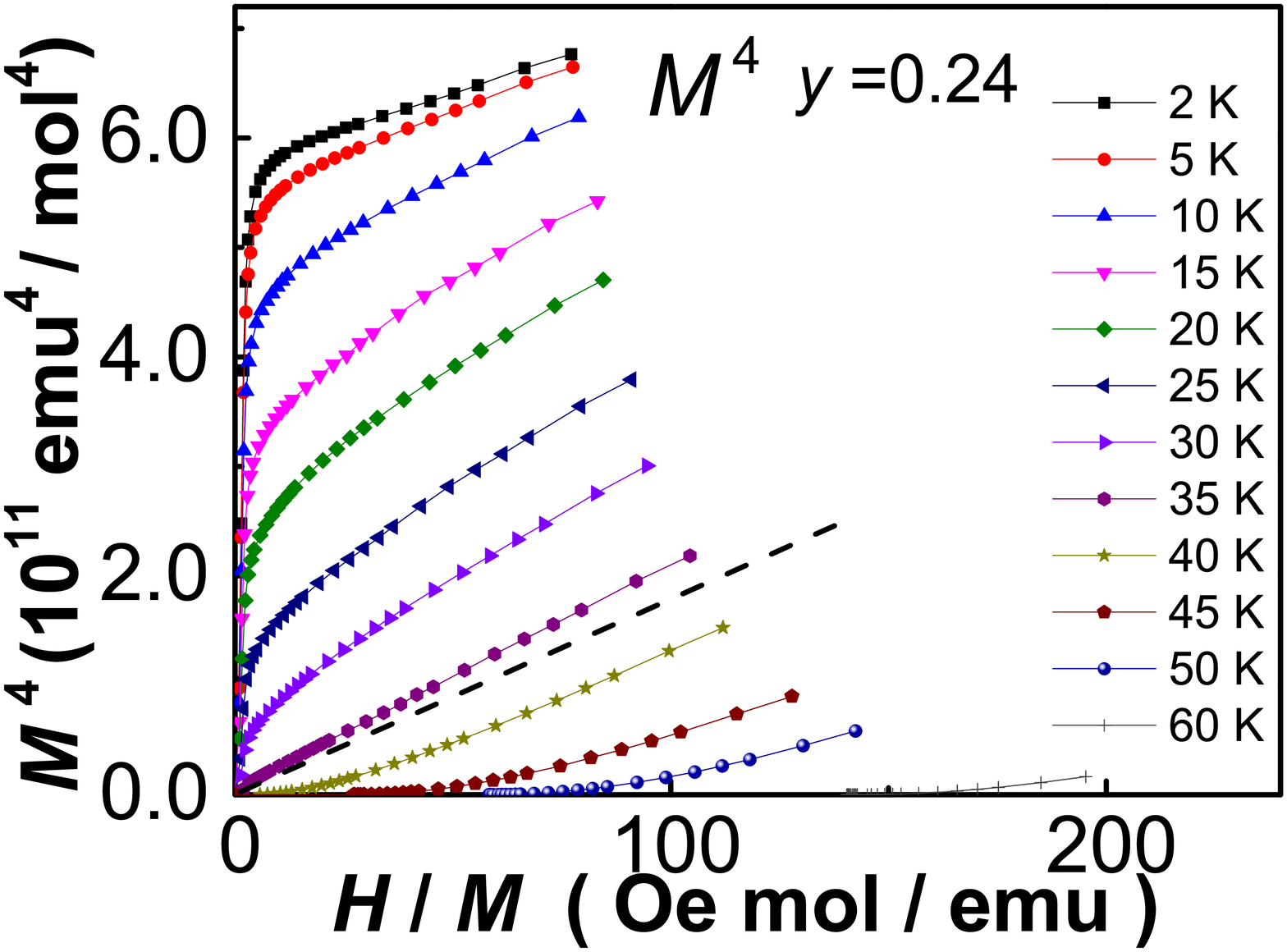}& \includegraphics[width=0.22\linewidth,keepaspectratio]{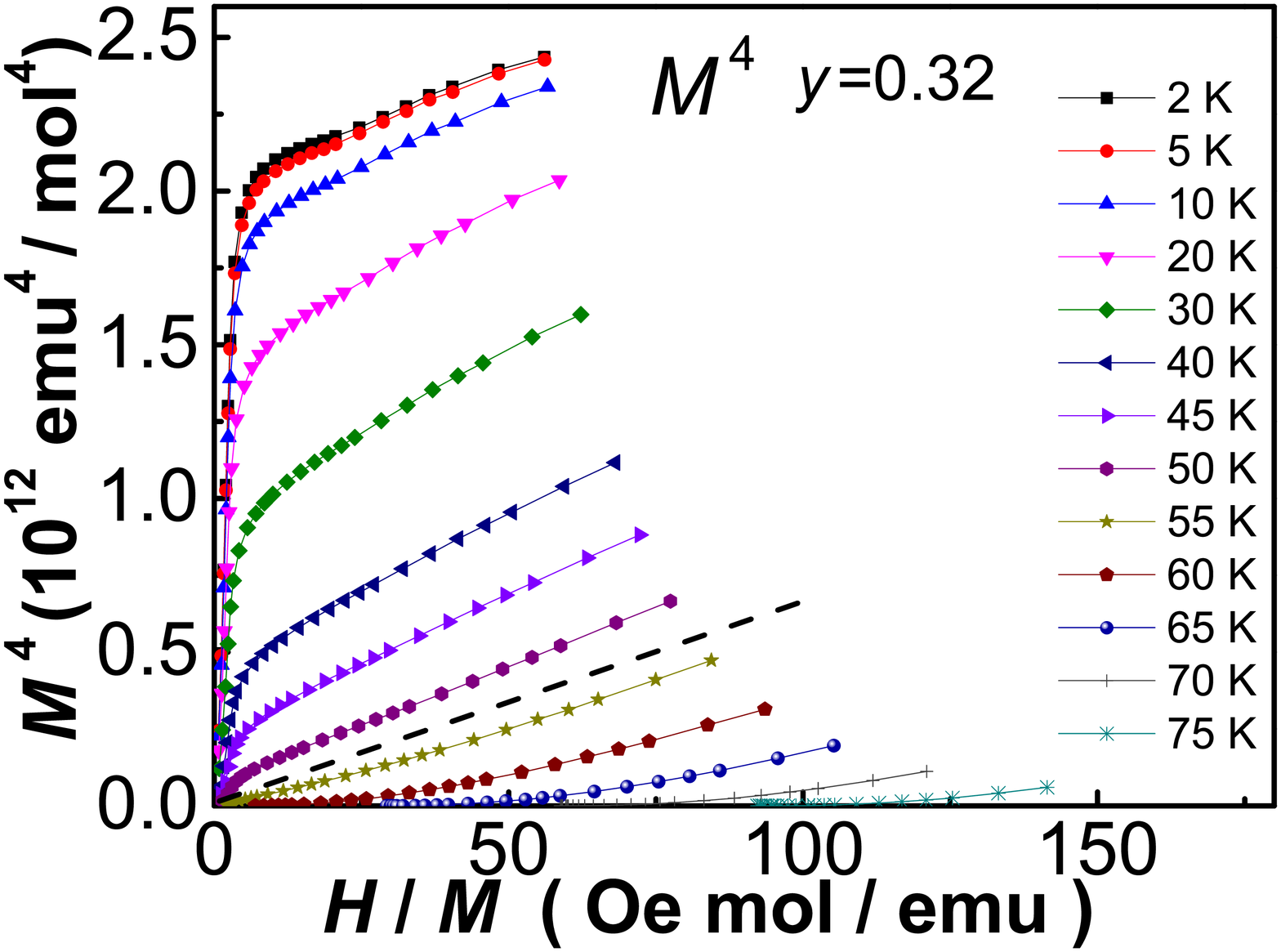}\\
	\end{tabular}
	\caption {
		\textbf{Arrott plots and \boldmath{$ M^4 $} plots.} 
		{\fontfamily{phvr}\selectfont
			$ M^2 $ versus $ H $ (Arrott plots) and $ M^4 $ versus $ H $ for FeGa$_{3-y}$Ge$_y$ with  $ y=0.14$, $0.15$, $0.16$, $0.18$, $0.24$, and $0.32$ respectively, as ({\fontfamily{phv}\selectfont a}) and ({\fontfamily{phv}\selectfont b}). Dash lines in ({\fontfamily{phv}\selectfont b}) are the description of the equation (\ref{eq:4}), and should be where $ M^4 $ plots shown up at the critical temperature $ T_{\mathrm{C}} $ (see text).  }
		\label{fig:M2-M4}}  
\end{figure*}

\begin{figure}[tbp] 
	\centering
	\begin{tabular}{cc}
		\includegraphics[width=0.4\linewidth,keepaspectratio]{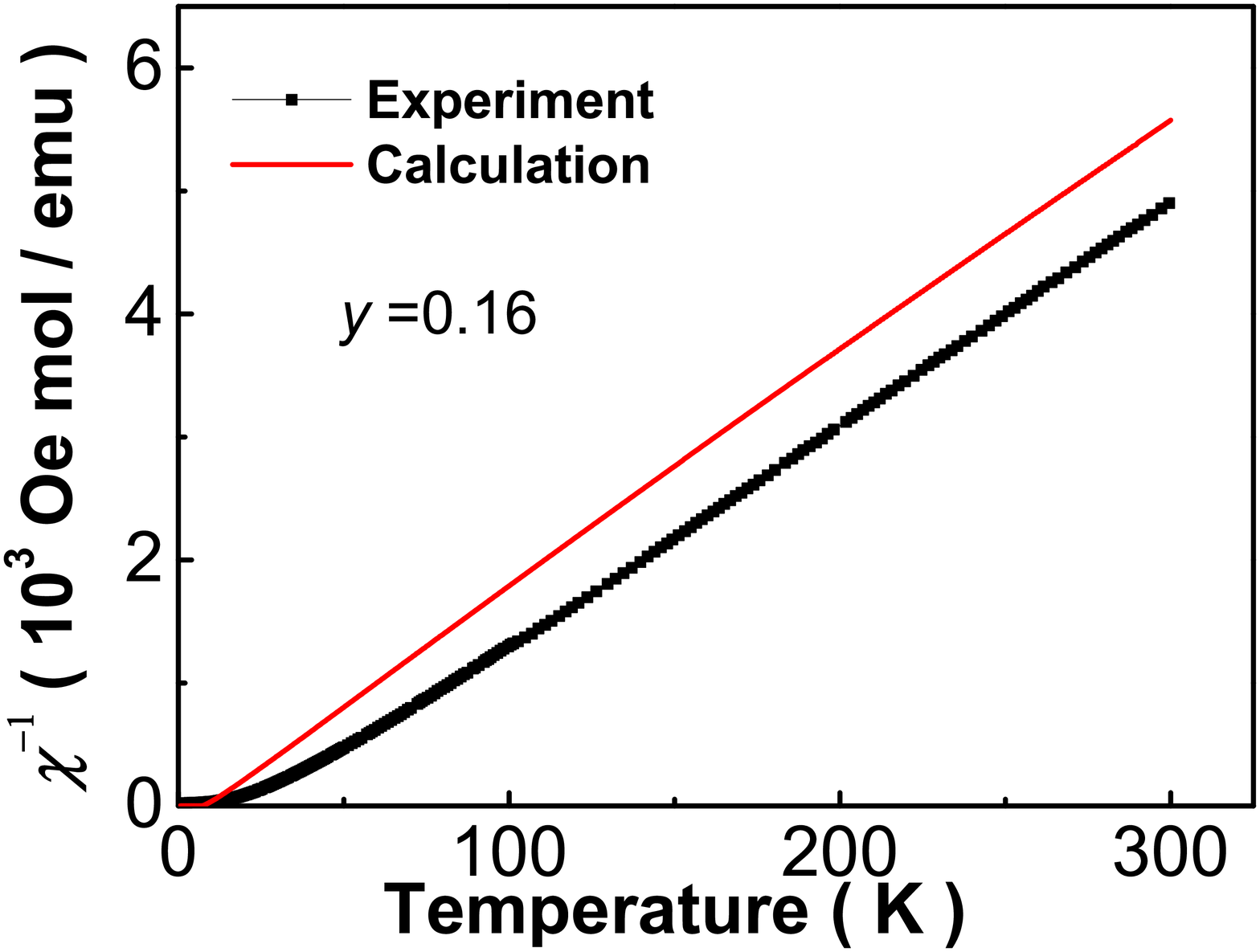}& \includegraphics[width=0.4\linewidth,keepaspectratio]{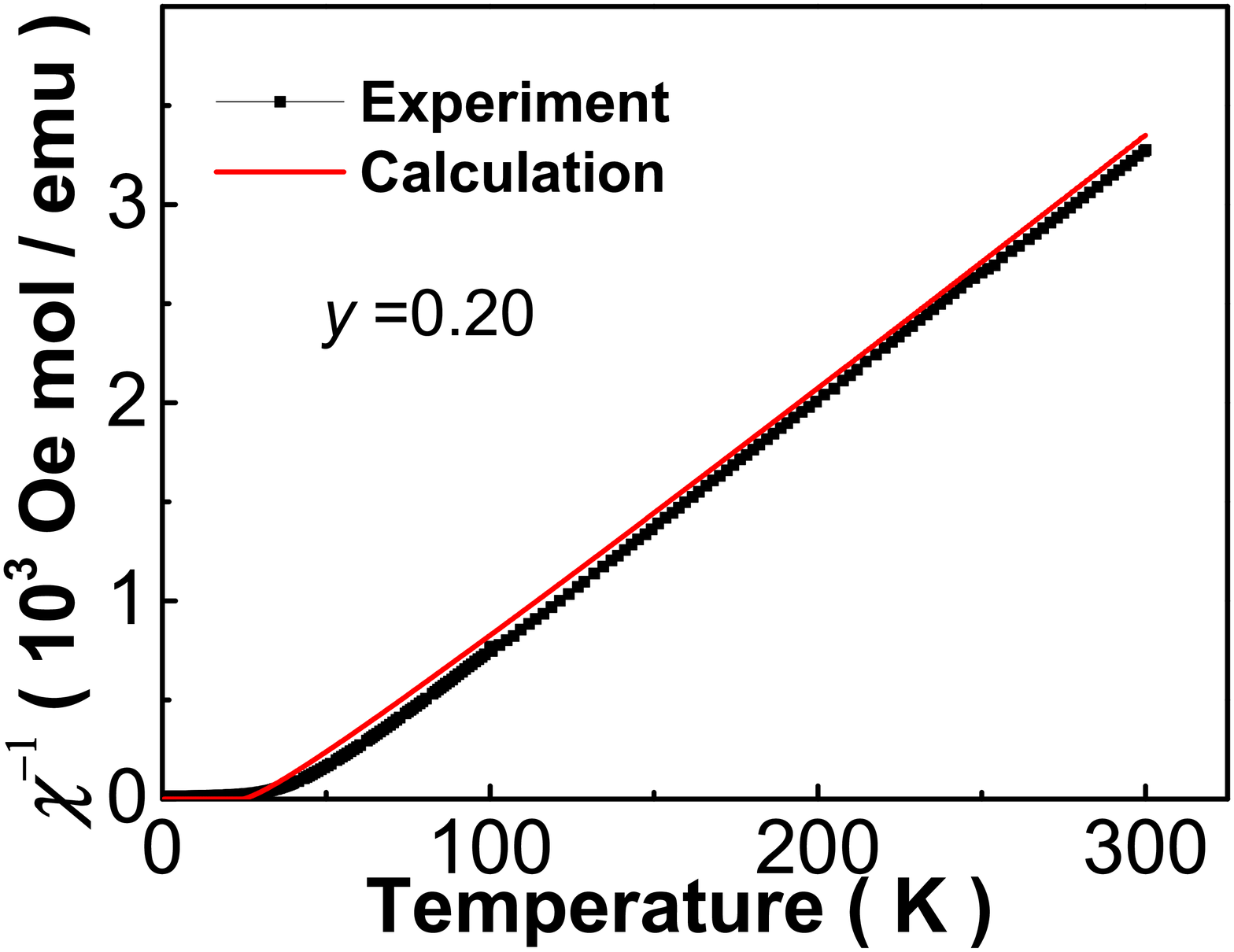}\\
		\includegraphics[width=0.4\linewidth,keepaspectratio]{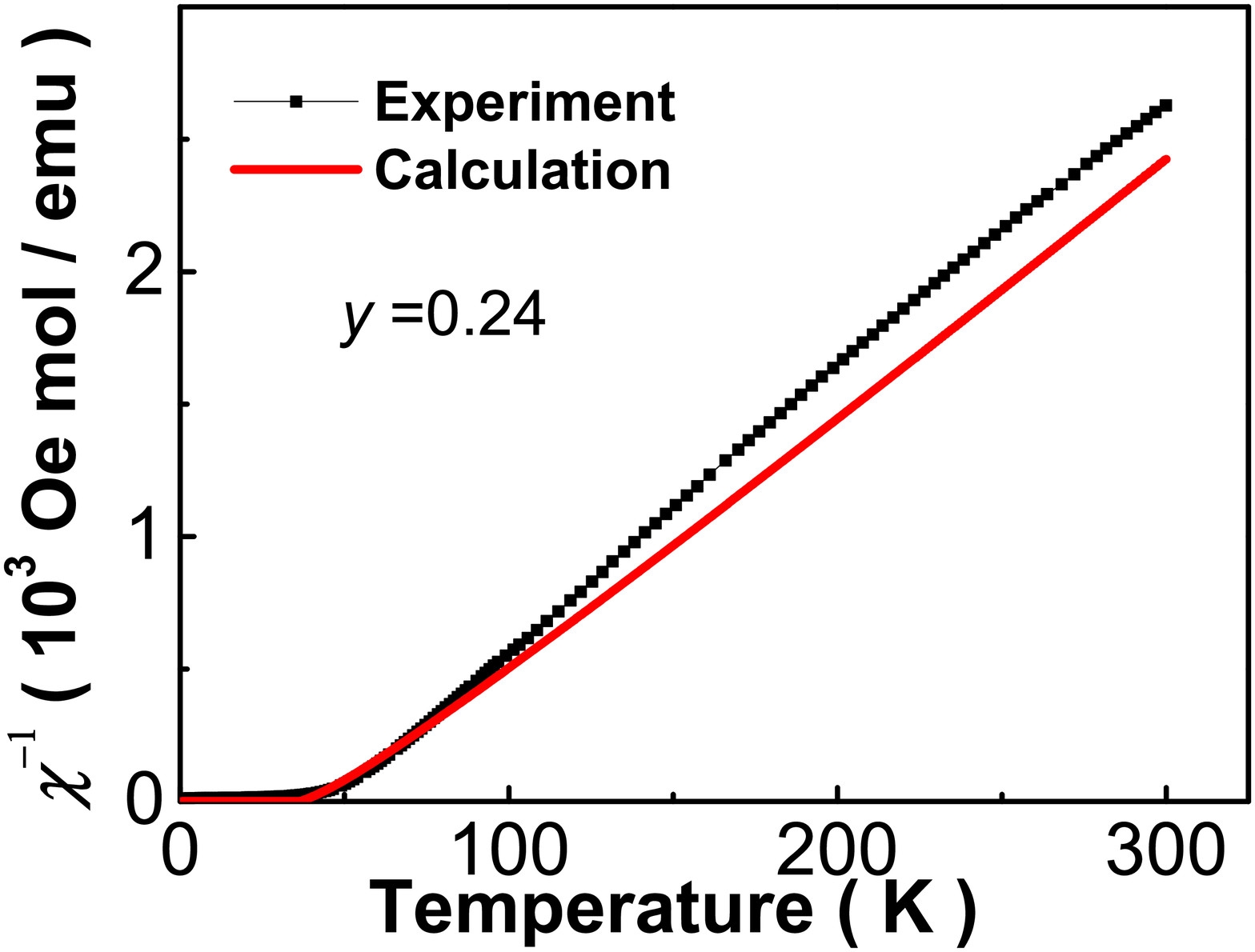}& \includegraphics[width=0.4\linewidth,keepaspectratio]{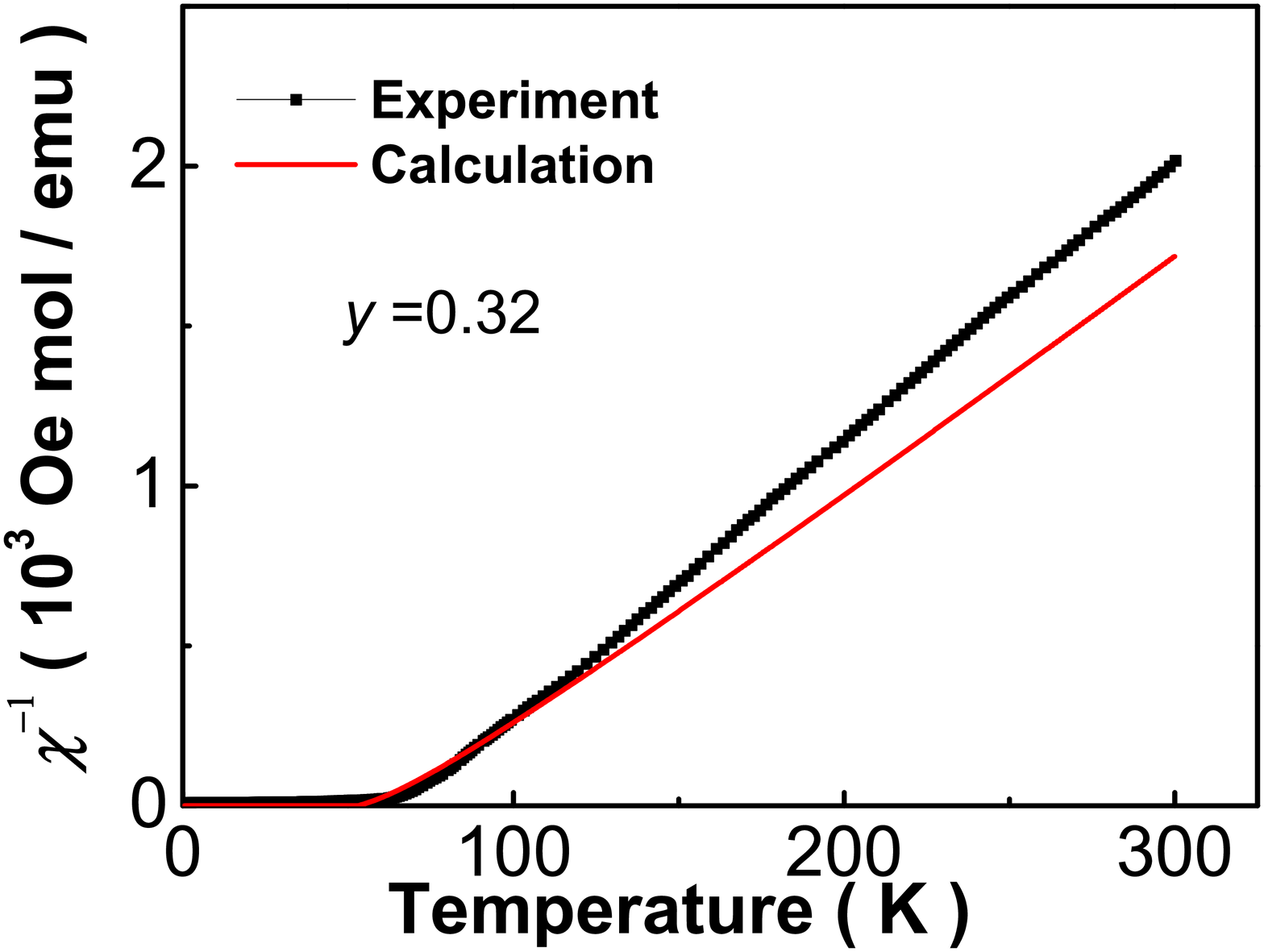}\\
	\end{tabular}
	\caption {\textbf{Temperature dependence of inverse susceptibility.} $ T $ dependences of $ \chi^{-1} $ for FeGa$_{3-y}$Ge$_y$ with $ y= 0.16 $, $ 0.20 $, $ 0.24 $ and $ 0.32 $. Black lines and squares represent experimental results. Red lines represent reconstructed results based on the theories of spin fluctuations (see text).
		\label{fig:CE}}  
\end{figure}

To investigate the character of the magnetic ordering, a Rhodes-Wohlfarth plot and a Generalized Rhodes-Wohlfarth plot (or so-called Deguchi-Takahashi plot developed by Takahashi \cite{Murata1972,Takahashi2013}) are drawn in Fig. \ref{fig:RW-DT}A and \ref{fig:RW-DT}B, respectively. $(1/2)P_{\mathrm{C}}$ in Fig. \ref{fig:RW-DT}A represents the effective spin per atom, whose value can be derived from $ {P_{\mathrm{eff}}}^2= P_{\mathrm{C}}(P_{\mathrm{C}}+2) $. As shown in Fig. \ref{fig:RW-DT}A, $ P_{\mathrm{C}}/P_{\mathrm{S}} $ of FeGa$_{3-y}$Ge$_y$ are not described by the fitting curve and have much smaller values than other ferromagnetic metals and alloys with the same $ T_{\mathrm{C}} $.  Unlike the majority of ferromagnetic metals or alloys, FeGa$_{3-y}$Ge$_y$ contains a considerably low effective Fermi energy caused by its sharp density of states at $ E_{\mathrm{F}} $ \cite{Arita2011,Gippius2014}, resulting in its $ T_{\mathrm{C}} $ to vary considerably less rapidly than the $E_{\mathrm{F}}$ and the failure to follow the Rhodes-Wohlfarth curve which well describes the behavior of most other metals and alloys \cite{Wohlfarth1961}. The largest value for  $ P_{\mathrm{C}}/P_{\mathrm{S}} $ obtained in this work is 2.6 at $y=0.16$ for FeGa$_{3-y}$Ge$_y$, corresponding to a weakly itinerant nature. The smallest $ P_{\mathrm{C}}/P_{\mathrm{S}} $ of 1.8 at $y=0.32$ indicates an adequate localized nature within the system, which is comparable with that of nickel (1.5). Also in Fig. \ref{fig:RW-DT}B, the magnitude of the magnetic ordering parameter $ T_\mathrm{C}/T_0 $ of FeGa$_{3-y}$Ge$_y$ is close to the localized regime value, 1, and spread towards the itinerant regime with increasing $y$, which is consistent with the results based on Fig. \ref{fig:RW-DT}a, suggessting a modulated state of magnetic moments is present in FeGa$_{3-y}$Ge$_y$.
      
Next, $ M^2 $ versus $ H/M $ plots (or so-called Arrott plots) and $ M^4 $ versus $ H/M $ plots are shown in Fig. \ref{fig:M2-M4}a (Fig. S4 and) and Fig. \ref{fig:M2-M4}b, (Fig. S5), respectively. According to the mean-effective-field solution of an arbitrary spin Ising model, if the Gaussian distribution of exchange coupling intensity is considerably greater than the mean value of exchange bonds, Arrott plots should show straight lines, and one plot must pass the origin at the critical temperature $ T_{\mathrm{C}} $. In the case of the FeGa$_{3-y}$Ge$_y$ system, only the Arrott plots for $ y = 0.32 $ show good linear behavior. The other samples show convex curvature even at $ T_{\mathrm{C}} $, and the curvature decreases with the increasing electron doping. In contrast, all the $ M^4 $ plots show good linear behavior, especially at $ T_{\mathrm{C}} $ where the $ M^4 $ plot passes the origin (0,0). Nonsignificant deviations from linear behavior for $ y=0.14 $ and 0.15 are observed in  $ M^4 $ plots (See Fig. \ref{fig:M2-M4}b and  Fig. S5). In these cases, the critical temperature $ T_{\mathrm{C}} $ can still be estimated by the low-field data of the isothermal Arrott plots approximating the arbitrary spin Ising model. $ T_{\mathrm{C}} $ reaches 0 at $y=0.15$, indicating the QCP, and as shown in Fig. \ref{fig:P-D}, FeGa$_{3-y}$Ge$_y$ with $y<0.15$ is paramagnetic, and samples with $y>0.15$ are weak ferromagnets.
  
 \textbf{Experiment vs. theory.} Experimental results of  $ \chi^{-1} $ versus $T$ and those of the theoretical reconstruction are shown in Fig. \ref{fig:CE} (and Fig. S6).
The reasonable consistency between experimental observations and theoretical calculations evidences the precision of the spin-fluctuation parameters we estimated in this work, and also the success of our analysis for the modulated ferromagnetic FeGa$_{3-y}$Ge$_y$.

\textbf{Universality of spin fluctuations.} In the generalized Rhodes-Wohlfarth plot shown in Fig. \ref{fig:RW-DT}b, the red line represents the generalized Rhodes-Wohlfarth theoretical equation $ P_{\textrm{eff}}/P_{\textrm{S}} =1.4\times(T_\mathrm{C}/T_0 )^{-2/3}  $, where $T_0$ represents the energy width of the dynamical spin fluctuation spectrum in frequency space corresponding to the stiffness of spin density in amplitude. 
FeGa$_{3-y}$Ge$_y$ with various amplitudes of dynamical spin fluctuations corresponded to different $T_0$ values as shown in Table 1 roughly satisfy the equation and are relatively widely spread along the line, with its $ T_{\textrm{C}} $ increasing from 0 at QCP to a considerably high value of 53.1 K, lying in the crossover region between the localized and the weakly itinerant regimes. The good fitting of the equation for the entire range of weak ferromagnets shown in Fig. \ref{fig:RW-DT}b implies a great reliance on spin fluctuations in reconciling the ferromagnets of different electron itinerancy from a localized regime to an itinerant regime.

\section{Discussion}

Magnetic behaviors that are intermediate between localized and itinerant nature in FeGa$_{3-y}$Ge$_y$ imply great difficulty in explaining the magnetic properties within a unified theory. Additionally, we should go beyond the models limited at the ground state in order to elucidate the temperature dependent magnetic properties involving the effects of spin fluctuations. Starting by dealing with the intrinsic free energy $ F $ in magnetization, which can be expanded in powers of magnetization $ M $ by tracking the splitting in band calculation:

\begin{equation}
F(M,T)=F(0,T)+\frac{1}{2}a_{1}(T)M^2 + \frac{1}{4}a_{2}(T)M^4+\cdots,
 \label{eq:1}
\end{equation}
Converted as the magnetic field $ H $ dependent $ M $ equation:
\begin{equation}
H =\frac{\partial F}{\partial M}=a_1(T)M+a_2(T)M^3+a_3(T)M^5+\cdots,  
 \label{eq:2}
\end{equation}
where  $ F(0,T) $ is the free energy at $ M=0 $, and  $ a_i (T) $ are expansion coefficients related with the electron density of states and its derivatives near $ E_{\mathrm{F}} $. 

The thermodynamic state of the free energy is determined by the association of the hopping conduction electrons with the repulsion by electrons with opposite spin directions on site. For an itinerant ferromangetic system, where its thermodynamic state becomes stable at finite mangentization, its magnetic properties can be described by the linear Arrott plot within coefficients $ a_1 $ and $ a_2 $ neglecting higher power terms, since the conduction electron density is fairly restricted around the Fermi energy $ E_{\mathrm{F}} $ in the ferromagnets, which leads to the famous equation:

\begin{equation}
M^2(M,T)=-\frac{a_1(T)}{a_2(T)}+\frac{1}{a_2(T)}\frac{H}{M(H,T)}.
 \label{eq:3}
\end{equation}

Numerous systems are governed by equation \ref{eq:3}, some weakly ferromagnetic compounds similar to FeGa$_{3-y}$Ge$_y$ have been reported as examples are ZrZn$ _{2} $ \cite{Ogawa1968}, Sc$ _{3} $In \cite{Takeuchi1979}, ZrTiZn${_2}$ \cite{Wohlfart1970}, ZrZn${_{1.9}}$ \cite{Wohlfart1970}, and Ni-Pt alloys \cite{Beille1974}. However, Arrott plots of ferromagnetic FeGa$_{3-y}$Ge$_y$ are not linear around the Curie points, especially when the positon, $y$, is close to the critical point of 0.15. This suggests the requirement for a higher power term of free energy $ a_3(T)M^5 $. The higher power term $ a_3(T)M^5 $ is not concerned by the ground-state-based theories such as Hartree-Fock approximation (HFA) or random-Phase approximation (RPA) etc. \cite{Izuyama1963,Bloch1929,Herring1966}. Even in the present form of the SCR theory, the fourth expansion coefficient, $a_2(T)$, is assumed to be temperature independent resulting in an inaccurate prediction that the spontaneous magnetic moment in ferromagets vanishes at Cuire temperature, this also implies the necessary for a higher power of term $ a_3(T)M^5 $ in the free-energy function for the approximation. Inputting all the dynamical parameters of $ a_i $ for the $ M^4 $ plots at the critical point $ T_{\mathrm{C}} $, we have \cite{Takahashi2013}:

\begin{equation}
H/M=\frac{T^{3}_{\mathrm{A}}}{2\mu_{\mathrm{{B}}}[3\pi T_{\mathrm{C}}(2+\sqrt{5})]^2}{\left( \frac{P_{\mathrm{S}}}{M_{\mathrm{S}}}\right) }^5M^4.
 \label{eq:4}
\end{equation}
where  spin-fluctuation parameter $T_{\mathrm{A}}$ represents the width of the distribution of the dynamical susceptibility in the wave vector space. For $ y= $ 0.14 and 0.15 in FeGa$_{3-y}$Ge$_y$, the small deviation from linear of the $ M^4 $ plots may be caused by the comparable $ a_2(T)M^3 $ terms and $ a_3(T)M^5 $ terms in the vicinity of the QCP, indicating the comparable effects in nonlinear couplings of spin fluctuations to the effects of non-negligible temperature dependence in general. For $y\geq 0.16$, the term $ a_3(T)M^4 $ gradually becomes overwhelming compared to $a_2(T)M^2$,  \Big($ \frac{T^{3}_{\mathrm{A}}\rho^3}{2\mu_{\mathrm{{B}}}\left(\rho'^2/\rho^2-\rho''/3\rho\right)[3\pi T_{\mathrm{C}}(2+\sqrt{5})]^2}{\left( \frac{P_{\mathrm{S}}}{M_{\mathrm{S}}}\right) }^5M^2\gg1$, where $\rho$ represents the density of states \Big), hence the $ M^4 $ plots show much better linear behaviors than the Arrott plots do, and synchronously the curvature begins to decrease in Arrott plots with the electron doping. For ferromagnetic FeGa$_{3-y}$Ge$_y$, we observe that the Arrott plots at $ T_{\mathrm{C}} $ nearly pass the origin, indicating the non-negligible temperature dependence of spin fluctuations is still considerable, even in the case where their $ T_{\mathrm{C}} $ approaches the minimum at the critical point.

The so-called Deguchi-Takahashi plots (Fig. \ref{fig:RW-DT}b) should also be considered in analyzing the doping effects in FeGa$_{3-y}$Ge$_y$, where the parameter, $T_{\mathrm{0}}$, characterizing the energy width of the dynamical spin fluctuation spectrum, is involved. In ferromagnets, if  $T_{\mathrm{0}}$  is comparable with $T_{\mathrm{C}}$ in magnitude, the localized nature of electrons becomes dominant according to the SCR theory of spin fluctuations and the related approximations \cite{Moriya1985,Takahashi1986,Takahashi2013}.The right side of the abscissa in Fig. \ref{fig:RW-DT}b, where $ T_{\mathrm{C}}/T_{\mathrm{0}} \sim 1 $, represents the extreme of localization, the left side where $ T_{\mathrm{C}}/T_{\mathrm{0}} \ll 1 $ represents the extreme of itinerancy. Fig. \ref{fig:RW-DT}b demonstrates that the nature of the electrons in FeGa$_{3-y}$Ge$_y$ is in the crossover region between localized picture and itinerant picture. In terms of the model of closed Kondo-Heisenberg approximation for a Kondo-Heisenberg lattice, the electron doping effects of the Kondo interaction $ \langle J_K\sum_iS_i\cdot s_{ci} \rangle $ become relatively weaker than the Heisenberg interaction $ \langle ({J_H}/z)\sum_{(ij)}S_iS_j\rangle $ does by the Ge substitution in the FeGa$_{3-y}$Ge$_y$ system, i.e., the itinerancy acquired from the Kondo effect in $ d $ electrons through intersite exchange becomes less significant by the continuous electron doping in  FeGa$_{3-y}$Ge$_y$. Importantly, the red line of the generalized Rhodes-Wohlfarth equation that fits the FeGa$_{3-y}$Ge$_y$ system in Fig. \ref{fig:RW-DT}B  fairly well describes not only the localized systems but also the itinerant ones by taking into account the effects of spin fluctuations, implying the crucial role of spin fluctuations in properly understanding ferromagnetism in systems showing coexistence of localized and itinerant characters.

Celebrated approximations such as HFA and RPA only deal with the paramagnetic contributions of spin fluctuations, however for FeGa$_{3-y}$Ge$_y$, effects of temperature dependent mode-mode coupling spin fluctuations on the thermal equilibrium state is crucial for its magnetic properties. We take the quantum statistical mechanical theory of SCR approximation of spin fluctuations into consideration, in which two well-known assumptions are inherited: (1) ~In the ground state, the magnetic properties can be described by the band calculation; (2) ~The effects of spin-spin couplings can be mainly represented by the second expansion coefficient of the free energy. We should mention that the theories of spin fluctuations are then in contrast with the phenomenological-theoretical-based technique of the Modified Arrott plot in which arbitrary critical exponents can be applied \cite{Moriya1985,Stanley1971}, since the function of free energy in the theories of spin fluctuation is even.

In the weakly ferromagnetic limit of the SCR approximation, the imaginary part of the dynamical spin susceptibility for ferromagnets is described by the double Lorentzian form in the small  $ q, \omega $-region \cite{Moriya1985}:

\begin{equation}
 \mathrm{Im}\chi(q,\omega)= \frac{\chi(0,0)}{1+q^{2}/\kappa^{2}} \frac{\omega \Gamma_{q}}{\omega^{2}+\Gamma_{q}^{2}}.
 \label{eq:5}
\end{equation}
where $\Gamma_q $ is the spectral width of the spin fluctuations given by $ \Gamma_q =(A/C)q(q^2+\kappa^2)=\Gamma_0q(q^2+\kappa^2) $, and $ \kappa^2=\varrho/2A\chi=N_0/2\bar{A}\chi $, and it leads to:

\begin{equation}
\frac{P_\mathrm{S}^2}{4}=\frac{15  T_0}{T_\mathrm{A}} c~ \left(  \frac{T_\mathrm{C}}{T_\mathrm{0}} \right)  ^{4/3},
  \label{eq:6}
\end{equation}
in weakly ferromagnetic systems.

Derived from equation \ref{eq:5}, the inverse magnetic susceptibility is given by \cite{Takahashi2013}:
\begin{equation}
y=\frac{N_0}{2T_A\eta^2}\dfrac{\kappa^2}{\kappa^2+q^2}\chi^{-1}\cong
\frac{\bar{F_1}P_\mathrm{s}^2}{8T_{\mathrm{A}}\eta^2}\left\lbrace -1+\frac{1+\nu y}{c}\int_0^{1/\eta}\mathrm{d}z z^3\left[ \mathrm{ln}u-\frac{1}{2u}-\Psi(u)\right]\right\rbrace.
  \label{eq:7}
\end{equation}
with $  u=z(y+z^2)/t $, $ t=T/T_{\mathrm{C}} $, $\nu=\eta^2T_{\mathrm{A}}/U$, $ \eta=(T_{\mathrm{C}}/ T_{\mathrm{0}})^{1/3} $, $ c=0.3353 $. $ \Psi(u) $ is the digamma function, and parameter $ \bar{F_1} $ is the mode-mode coupling constant, representing the fourth order expansion coefficients of magnetic free energy. $ \bar{F_1}=N_A^3(2\mu_{\mathrm{ B}})^4/\zeta k_{\mathrm{B}}$, $ \zeta $ is the slope of the Arrott plots at low temperature, $ N_\mathrm{A} $ and $ k_{\mathrm{B}} $ are Avogadro's number and Boltzmann constant \cite{Takahashi1985}. 

Due to the compensation of the increasing thermal amplitude of spin fluctuation for the suppression of the zero-point spin fluctuation under applied magnetic field with increasing temperature, the local spin amplitude squared at finite temperature can be treated as nearly conserved, which leads to \cite{Takahashi1986}:

\begin{equation}
\bar{F_1}=\frac{4}{15}\frac{k_{\mathrm{B}}T_\mathrm{A}^2}{T_0}.
  \label{eq:8}
\end{equation}

Then all the spin-fluctuation parameters can be estimated merely from the macroscopic magnetization measurements, without the need of pursuing any extra dynamical measurements such as nuclear magnetic resonance or neutron scattering  \cite{Ishikawa1985,Bernhoeft1983,Corti2007,Yoshimura1988,Yoshimura1999,Yoshimura1987}.

The quantitative agreement between the theoretical reconstruction and experimental results shown in Fig. \ref{fig:CE} (and Fig. S6) implies the success of the elucidation of the magnetization that involves the spin fluctuations for the intermediate FeGa$_{3-y}$Ge$_y$ system. Moreover, our analysis successfully explains the ferromagnetic FeGa$_{3-y}$Ge$_y$ ranging from the adequate localized region to the itinerant regime that well fits the generalized Rhodes-Wohlfarth relation $ P_{\mathrm{eff}}/P_{\mathrm{S}}  =1.4\times(T_\mathrm{C}/T_0 )^{-2/3}  $ describing various ferromagnets as shown in Fig. \ref{fig:RW-DT}B, indicates a potential universality to quantitatively explain magnetism in weakly ferromagnetic systems in a broad $ T_{\mathrm{C}} $ range.

In summary, we have shown that electron doping by Ge substitution substantially affects the magnetic ground state and spin-spin correlation in FeGa$_{3-y}$Ge$_y$, causing phase transitions as well as changes in magnetic orderings within the system. We successfully take the temperature-dependent effects of spin fluctuations in general into account for the modulated ferromagnetic FeGa$_{3-y}$Ge$_y$, and the theoretical results agree well with the experimental observations. Our analysis shows a potential universality for the entire range of weakly itinerant ferromagnetic systems by involving the spin fluctuations. FeGa$_{3-y}$Ge$_y$ should be the promising model system to unify the magnetic theory for localized and itinerant electrons.

\section{Methods}

\textbf{Experiment.} 
Single crystals of  FeGa$_{3-y}$Ge$_y$ were synthesized by Ga self-flux method. Powders of Fe (99.99\%), Ge (99.99\%) and Ga (99.9999\%) ingot with the ratio of Fe : Ge : Ga = 1 : $ Y $ : 9  $ (0.01\leq Y \leq 3) $ were loaded and sealed in an clean evacuated silica tube. The mixture were melted and homogenized in a furnace at 1273 K for 40 hours, and cooled to room temperature slowly. Excess Ga flux was removed with aqueous solution of H$ _2 $O$ _2 $ and HCl. X-ray diffraction pattern confirmed the samples are single crystal in FeGa$_3$ type structure without second phase. The chemical composition of FeGa$_{3-y}$Ge$_y$ was determined by wavelength-dispersive electron microprobe analysis (EPMA). Since Ge is insoluble over entire range, the maximum $ y $ obtained in this work is 0.32. The lattice parameter $ a $ increases from 6.263 to 6.279 \AA{} and $ c $ decreases from 6.554 to 6.540 \AA{} with Ge substitution, following Vegard's law, which is consistent with the former report \cite{Umeo2012}. The magnetization $ (M ) $ of FeGa$_{3-y}$Ge$_y$ was measured as a function of $ T $ and $ H $ up to 7 T by the superconducting quantum interference device (SQUID) magnetometer in Research Center for Low Temperature and Materials Science, Kyoto University. The electrical resistivity measurements were employed on a home-built quadrupole electrical conductivity measuring device from 5 to 300 K.

\section{Acknowledgments}
\begin{acknowledgments}
We thank Y. Takahashi for commenting on the manuscript and useful discussions. This work is supported by Grants-in-Aid for Scientific Research 22350029 and 26410089 from the Ministry of Education, Culture, Sports, Science and Technology of Japan and Grants for Excellent Graduate Schools, MEXT, Japan.
\end{acknowledgments}

\clearpage 

\clearpage 

\vspace{25mm}

\clearpage
\begin{table}[tp]
	\begin{center}
		\begin{tabular*}{\linewidth}{@{\extracolsep{\fill} }lccccccc} \hline\hline
			$ y $ & $ P_{\mathrm{eff}} $ & $ P_{\mathrm{C}} $ & $ P_{\mathrm{S}} $ & $ T_{\mathrm{C}} $ & $ T_{\mathrm{A}} (10^4) $ & $ T_{\mathrm{0}} (10^2)$ & $ \bar{F_1} (10^5) $ \\ \hline
			0.16 & 0.71 & 0.226 & 0.087 & 7.2  & 7.56 & 1.10 & 1.39 \\
			0.18 & 0.74 & 0.244 & 0.112 & 14.3 & 9.67 & 1.87 & 1.33 \\
			0.20 & 0.79 & 0.274 & 0.133 & 24.8 & 1.23 & 2.99 & 1.35 \\
			0.21 & 0.80 & 0.281 & 0.136 & 32.6 & 1.42 & 3.48 & 1.59 \\
			0.24 & 0.90 & 0.345 & 0.156 & 36.4 & 1.33 & 4.23 & 1.11 \\
			0.27 & 0.91 & 0.352 & 0.187 & 46.9 & 1.29 & 4.30 & 1.03 \\
			0.32 & 0.96 & 0.386 & 0.216 & 53.1 & 1.18 & 4.56 & 0.73 \\ \hline\hline
		\end{tabular*}
	\end{center}
	\caption{ Spin-fluctuation parameters estimated from Arrott plot and $ M^4 $ plot for $ y = 0.16, 0.18, 0.20, 0.21, 0.24, 0.27, $ and $ 0.32 $. $ P_{\mathrm{eff}} ,  P_{\mathrm{S}}, T_{\mathrm{C}}, T_{\mathrm{A}}, T_{\mathrm{0}} $ , and $ \bar{F_1} $, represent effective magnetic moment ( $ \mu_{\mathrm{B}} $/Fe atom), spontaneous magnetic moment at ground state ( $ \mu_{\mathrm{B}}$/Fe atom), Curie temperature (K), the width of the distribution of the dynamical susceptibility in the $ q $ space (K), the energy width of the dynamical spin fluctuation spectrum (K), and fourth order expansion coefficients of magnetic free energy (K), respectively. $ \frac{1}{2}P_{\mathrm{C}}$ represents effective spin per atom ( $ \mu_{\mathrm{B}}$ ).}
	\label{tab:1}  
\end{table}

\end{document}